\documentclass[11pt,a4paper]{article}
\usepackage{amsmath}
\usepackage{amsfonts}
\usepackage{amssymb}
\usepackage{graphicx}

\newtheorem{axiom}{Theorem}[section]

\newtheorem{guess}{Remark}[section]

\newtheorem{proposition}{Proposition}[section]
\newtheorem{definition}{Definition}[section]

\begin{document}
\bibliographystyle{unsrt}
\def\boxit#1#2{\setbox1=\hbox{\kern#1{#2}\kern#1}%
\dimen1=\ht1 \advance\dimen1 by #1 \dimen2=\dp1 \advance\dimen2 by
#1
\setbox1=\hbox{\vrule height\dimen1 depth\dimen2\box1\vrule}%
\setbox1=\vbox{\hrule\box1\hrule}%
\advance\dimen1 by .4pt \ht1=\dimen1 \advance\dimen2 by .4pt
\dp1=\dimen2 \box1\relax}

\def\build#1_#2^#3{\mathrel{\mathop{\kern 0pt#1}\limits_{#2}^{#3}}}

\def\K{\Bbb K}
\def\C{\Bbb C}
\def\R{\Bbb R}
\def\N{\Bbb N}
\def\ecart{\noalign{\medskip}}
\font\tenmsb=msbm10 \font\sevenmsb=msbm7 \font\fivemsb=msbm5
\newfam\msbfam
\textfont\msbfam=\tenmsb \scriptfont\msbfam=\sevenmsb
\scriptscriptfont\msbfam=\fivemsb
\def\Bbb#1{{\fam\msbfam\relax#1}}
\vfill\eject
 
\title{Local contractivity   of the $\Phi_0^4$
mapping}
\author{
 by\ \ Marietta Manolessou\\
EISTI\, \, -\, \, Departement of Mathematics}
\maketitle

\begin{abstract}

Previous results about the non trivial solution of the
$\Phi^4$-equations of motion for the Green's functions
 in the Euclidean\  space (of  $0\leq r\leq 4$ dimensions)
 in the Wightman Quantum Field theory framework, are reviewed
in the $0$ - dimensional case from the following three aspects:
\begin{itemize}
\item{}

The structure of the subset $\Phi \subset {\cal B}$
 characterized by the signs and ``splitting'' (factorization properties)
 is reffined and more explictly described by a subset
 $\Phi_0\subset \Phi$ 
    \item{}
 The local contractivity of the corresponding $\Phi ^4_0$ mappping is established
in a neighborhood   of a precise nontrivial sequence $H_0\in  \Phi_0$
 using a new norm in the Banach space $ {\cal B}$.
\item{} A new $\Phi ^4_0$  iteration is defined in the neighborhood of this sequence
$H_0\in  \Phi_0 $  and our numerical study displays 
clearly the stability of $\Phi_0$ (the splitting, the  bounds and sign
properties are perfectly illustrated),
 and a rapid convergence to the unique fixed point  $H^* \in \Phi_0$(cf. \cite{MMST}).

\end{itemize}
\end{abstract}
\vfill\eject

\vfill\eject

\section{ Introduction}
\pagenumbering{arabic}

\subsection{A new non perturbative method}

Several years ago we started a program for
 the construction of a non
 trivial $\Phi^4_4$  model consistent with the
general principles of a Wightman Quantum Field
 Theory
($Q.F.T.$) \cite{(Q.F.T.)}. In references \cite{MM1}, \cite{MM2} we have
introduced
 a non perturbative method for the
 construction of a non trivial solution of
 the system
 of the $\Phi^4$ equations of motion
for the Green's functions, in the Euclidean space
 of zero, one, two and four dimensions.

This method is different in approach from the work
 done in the Constructive $Q.F.T.$ framework of
 Glimm-Jaffe and others \cite{G.J.}, and the methods
 of Symanzik who created the basis for a pure
 Euclidean approach to $Q.F.T.$ \cite{Sym}.

    In the $Q.F.T.$ language the interaction of
 four scalar fields $\Phi(x)$ is described
 mathematically
 in the four-dimensional Minkowski space
with coordinates:
\begin{equation}\ x=\lbrace\   \ \vec x\in
\Bbb R ^3, x_0\in \Bbb R;
 ||x||=\sqrt{ [x_0^2 -\vec x ^2 ]} \rbrace
\label{(1.1)}
\end{equation}
by the following two-fold set of dynamical equations:
    \begin{itemize}
\item[i.] a nonlinear differential equation (the equation of
motion) resulting from the corresponding Lagrangian by application
of
 the variational principle :
    \begin{equation}    - (\boxit{1pt}{${ \ }^{ }$} + m^2 )
\Phi(x) =\frac {\Lambda}{\rho +\gamma}
 \lbrack\,:\Phi^{3}(x): - \alpha\Phi(x)\, \rbrack
\label{(1.2)}
    \end{equation}
 \item[ii.] the ``conditions of quantization''
 of the field $\Phi(x)$ expressed by the
 commutation relations:
            \begin{equation}[\Phi(x), \Phi(y)]  =
 [\dot {\Phi}(x), \dot {\Phi}(y)] = 0
\label{(1.3)}
\end{equation}
    \begin{equation}    [\Phi(x),\dot {\Phi}(y)]  =
 i{\gamma\over\rho +\gamma}
  \delta^3(\vec x - \vec y),\,x_0=y_0
\label{(1.4)}
\end{equation}
\end{itemize}

Here $m>0$  and $\Lambda>0$ are the physical mass
 and coupling constant of the interaction model,
and $\alpha$, $\beta$, $\gamma$, are physically
 well defined quantities associated to this model,
the so called renormalization constants.
 For the precise definition of the latter
 and of the normal
product  $:\Phi^{3}(x) :$ we refer the reader
 to the references \cite{Zim1}, \cite{MM3}.

From these equations one can formally derive an equivalent
infinite system of non linear integral
 equations of motion
for the Green's functions (the {\it ``vacuum expectation
values''}) of the theory
 (analogous, but not identical, to the Dyson
-Schwinger equations \cite{Dys}\cite{Sch}). This dynamical system
has been established in $4$ dimensions
  by using the Renormalized Normal Product
of \cite{MM3}.

The method is based on the proof of the existence and uniqueness
of the solution of the corresponding infinite system of dynamical equations
 of motion satisfied by the sequence of the
Schwinger functions, i.e. the connected, completely
 amputated with respect to the free propagator
Green's functions: $ H =\{H^{n+1}\}_{n =2k+1, k \in\Bbb N}$,
 in the Euclidean $r$-dimensional
 momentum space, ${\cal E}^{rn}_{(q)}$
 (where $ 0\leq r\leq 4 $).

 Notice that in the previous notations
 and in  what follows $\Bbb N$ means the set
 of non negative integers and $n$ will always be
 an odd positive
 integer.

\subsubsection{The equations of motion for the Schwinger functions and the $\Phi$-Iteration}

The infinite system of equations for the Schwinger
 functions derived by the system
\ref{(1.1)}, \ref{(1.2)}, \ref{(1.3)}, \ref{(1.4)}  has precisely
the following form in 4 dimensions:
    \begin{equation}
    H^2 (q,\Lambda)= -{\Lambda\over{\gamma+\rho}}
\lbrace\lbrack N^{(3)}_3H^4\rbrack
 -\Lambda\alpha
H^2 (q,\Lambda)\Delta_F(q) \rbrace
+{(q^2+m^2)\gamma\over{\gamma+\rho}}
    \label{1.10}
 \end{equation}
and
$$\forall\,  n \geq3, \, (q,\Lambda) \in
{\cal E}^{4n}_{(q)}\times \Bbb R$$
 \begin{equation}H^{n+1}(q,\Lambda) =  {1\over{\gamma+\rho}} \lbrace\,
 \lbrack A^{n+1}+ B^{n+1}
 + C^{n+1}\rbrack(q,\Lambda) + \Lambda
\alpha  H^{n+1}(q,\Lambda)\Delta_F(q) \rbrace
   \label{1.11}
 \end{equation}
with:
\begin{equation}
A^{n+1}(q,\Lambda) = - \Lambda\lbrack
N^{(n+2)}_3H^{n+3}\rbrack(q,\Lambda);
\end{equation}
\begin{equation}B^{n+1}(q,\Lambda) = - 3\Lambda\sum_{\varpi_n(J)}
\lbrack N^{(j_2)}_2H^{j_{2}+2}\rbrack
\lbrack N^{(j_1)}_1H^{j_{1}+1}\rbrack(q,\Lambda); \end{equation}
\begin{equation}C^{n+1}(q,\Lambda) = - 6\Lambda\sum_{\varpi_n(I)}\prod_{l=1,2,3}
\lbrack N^{(i_l)}_1H^{i_{l}+1}\rbrack (q_{i_{l}},\Lambda);
\end{equation}

Here the notations:

  $\lbrack N^{(n+2)}_3H^{n+3}\rbrack$ , $\lbrack N^{(j_2)}_2H^{j_{2}+2}\rbrack
\lbrack N^{(j_1)}_1H^{j_{1}+1}\rbrack$ and  $\displaystyle\prod_{l=1,2,3}
\lbrack N^{(i_l)}_1H^{i_{l}+1} \rbrack$, represent the  $\Phi^4_4$
operations which have been introduced in the
 \emph{ Renormalized
 \,  G-Convolution\ \ Product} $(R.G.C.P) $  context of the references
 \cite{Br.MM}
\cite{MM4}.
  Briefly, the two
 loop  $\Phi^4_4$ - operation is defined by:
 \begin{equation}
\lbrack N^{(n+2)}_3H^{n+3}\rbrack   = \int R^{(3)}_G \lbrack\
 H^{n+3}\prod_{i=1,2,3}
\Delta_F(l_i)\ \rbrack d^{4}k_1d^{4}k_2
\end{equation}

\begin{figure}[h]
\begin{center}
\hspace*{-5mm}
 \includegraphics[width=10cm]{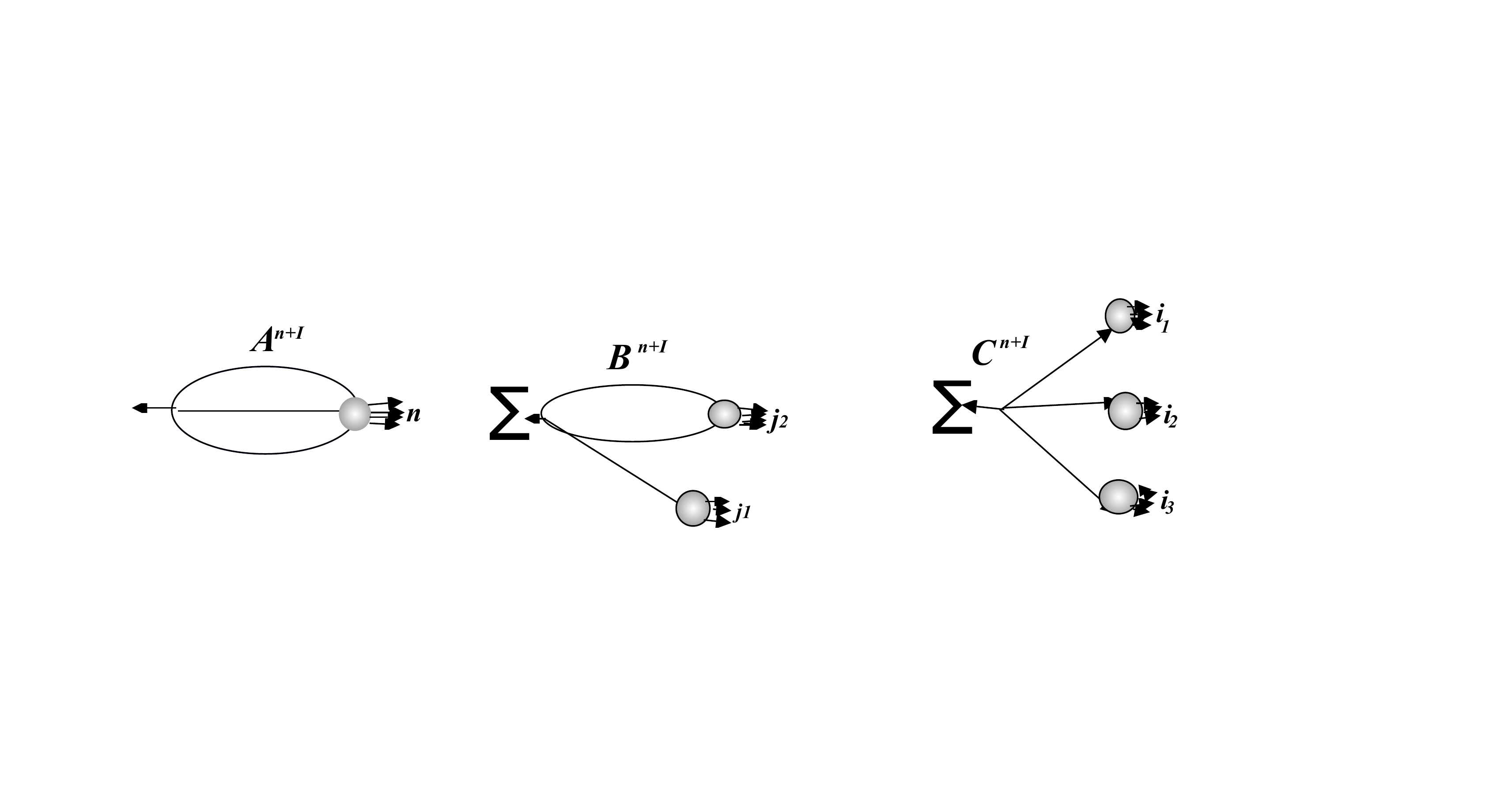}
\end{center} 
 \caption{\small\textrm{Graphical representation of the global terms of the $\Phi^4$ equations of motion}}
\label{fig.1.2}
\end{figure}

with $ R^{(3)}_G$ the corresponding renormalization operator
 for the two loop graph.

The analogous expression for the one loop  $\Phi^4_4$ -
 operation is the following:

\begin{equation}
\lbrack N^{(j_2)}_2H^{j_{2}+2}\rbrack
\lbrack N^{(j_1)}_1H^{j_{1}+1}\rbrack=
H^{j_{1}+1}(q_{j_1}) \Delta_F(q_{j_1})\, \int R^{(2)}_G \lbrack
H^{j_{2}+2}\prod_{i=1,2}\Delta_F(l_i)\rbrack d^{4}k
\end{equation}

with $R^{(2)}_G$, the corresponding renormalization  operator for
the
 two loop graph.

The notation $\Delta_F$ indicates the free propagator and the
$\Phi^4_4$
  operation
$N^{(j_1)}_1$ is exactly the multiplication (\emph{``trivial
convolution''})
 by  the corresponding
 free propagator $\displaystyle\Delta_F ={ 1\over (\|q\|^2+m^2)}$.
 Here $\|q\|^2$
 means the
 Euclidean norm of the vector $q\in{\cal E}^{4n}_{(q)}$.

    The information concerning the special features
 of the dynamics of four interacting fields,
 has been obtained through an iteration of these
 integral equations of motion in the two dimensional
 case,
 at fixed coupling constant and at zero external momenta,
 with the free solution as starting point.
 This is what was called the \emph{``$\Phi$-Iteration''}
 in \cite{MM1}.
 The exploration of the detailed structure of the
\emph{``$\Phi$-Iteration''}, has brought forth
 the particular properties of the different global terms
 of the equations at every order $\nu$,
 constructed in terms of the $H^{n+1}_{\nu}$ functions.
These properties essentially  were:
 \begin{itemize}
 \item{(a)} \emph{alternating signs}  and \emph{splitting}
 (or \emph{factorization})
 properties at zero external momenta:
\begin{equation}    H^{n+1}(q=0,\Lambda)=
 - n(n-1)\delta_n(\Lambda)
H^{n-1}(q=0,\Lambda) [H^2(q=0,\Lambda)]^2
\end{equation}
with $\{\delta_n\}$ a bounded increasing sequence of continuous
functions of $\Lambda$ and uniformly convergent to some finite
positive, $\delta_{\infty}$.

\item{(b)} \emph{ bounds at zero external momenta} and \emph{zero
momentum dominance
 bounds}, which in turn yield  global bounds
 of the general  form:
                    \begin{equation}
H^{n+1}(q,\Lambda) \leq  n\ !\   \  K^n
\end{equation}
 where $K$ is a finite positive constant independent of $n$.

  These features formed a self-consistent
 system of conditions conserved by the
\emph{``$\Phi$-Iteration''}.
 In particular they implied
 precise \emph{ ``norms''}
 of the sequences of the Green's functions $H^{n+1}$:
\begin{equation}
\| H \| = \sup_{q, n, \Lambda}\{ M_n^{-1}(q, \Lambda)
 \ \ | H^{n+1}(q,\Lambda)| \}
\label{1.7}
\end{equation}

with the  corresponding \emph{ norm\  \ functions} given as
follows:

$\forall\quad (q,\Lambda) \in {\cal E}^{4n}_{(q)}\times \Bbb
R^{+*} $
\begin{equation}M_n (q, \Lambda) = n !\ \
 (\ 1/8\ )^{n-5\over2}(\delta_{\infty})^
{n-1\over2}\ \ \prod_{i=1}^{n}\ \ M_1(q_i, \Lambda), \label{1.8}
\end{equation}
    with:
\begin{equation}M_1 (\ q_i,\Lambda)=
\lbrack\|q_i\|^2+m^2\rbrack^{1+{1\over8}}+10\Lambda^2 \label{1.9}
\end{equation}
and $\delta_{\infty}$ a universal constant which
characterizes the splitting properties presented previously
\end{itemize}

 These norms in turn are conserved and
automatically ensure
 the convergence of the
  \emph{``$\Phi$-Iteration''} to the solution.
 So, the answer
to the problem of how to construct the fixed-point method became
clear: it would be sufficient first
 to define a Banach space $\cal B$
 using the norms provided by the \emph{``$\Phi$-Iteration''}.
 One then should seek a fixed point of
 the equations of motion inside a characteristic
 subset $\Phi\subset \cal B$
  which exactly \emph{imitates}
 the fine structure of the $\Phi$-Iteration.

 This is exactly what we  tried to realize  for  $0\leq r\leq 4$
 dimensions in the references \cite{MM1} and \cite{MM2}.
We notice that, even the zero-dimensional
 corresponding system includes
 already the supplementary or
 \emph{``renormalization''} conditions
 imposed on the Schwinger functions
 in order to avoid the infinities
 appearing in the convolutions of the
 larger dimensions.

 These renormalization
 constraints constitute
 the crucial two-fold advantageous
 aspect of our method. On one hand
 they provide us with a non
trivial solution of the equations of motion even
 in four dimensions,
 a basic difference between our
approach and the \emph{``Constructive\
 \ Field\  \ Theory''} methods
 \cite{G.J.1}\cite{G.J.S.},
 which use the generating
functional formalism, in order to construct  the $Q.F.T.$ Green's
functions-moments of this
 functional.
 
On the other hand, this is the essential
 point that makes our system
 of equations in zero dimensions
equivalent to that obtained directly by derivation of the
generating
 functional
(application of the variational principle).

 This last point is precisely
 the argument that we used
in reference \cite{MM4} for the verification of the positivity condition
of the zero-dimensional solution.

\subsubsection{The ``new mapping''}

Using the \emph{``$\Phi$-Iteration''} which had the free solution
 as starting point
 we discovered
  (in two dimensions, at zero external momenta) at every order $\nu$,
 the crucial
properties
 presented in the previous section,
 of the global terms $A^{n+1}$, $B^{n+1}$, $C^{n+1}$, (constructed
 in terms of the $H^{n+1}_{(\nu)}$ functions), namely:
 \emph {``alternating signs''}, \emph{``splitting''}
(or \emph{factorization})
 properties, \emph{``bounds at zero external momenta'' and
``zero momentum dominance bounds''},
 which in turn yield  global bounds and conserved norms.

As we pointed out in the previous section these ``conserved
norms''(1.1.8) lead to the convergence of the \emph{
``$\Phi$-Iteration''}. So, by
 introducing an appropriate Banach space ${\cal B}$ defined by
 these norms and a characteristic subset $\Phi\subset {\cal B}$, 
 which exactly
imitates the fine structure of the \emph{ ``$\Phi$-Iteration''}, 
 one expects
to establish
 by a fixed point theorem the existence of a non trivial solution
 inside this subset.

    Unfortunately this is not the case. The  global terms $A^{n+1}$,
 $B^{n+1}$,
and $C^{n+1}$,
 (tree terms) (with alternating signs)
 have the same asymptotic behaviour with respect to $n$.
 More precisely,
at every fixed value of the external momenta,we obtain:

\begin{equation} A^{n+1}\build\sim_{n\rightarrow\infty}^{}
(-\ \delta_{\infty})^{n-1\over2}
  n !\  n^2 \end{equation}
    \begin{equation}B^{n+1}\build\sim_{n\rightarrow\infty}^{}
 -  (-\ \delta_{\infty} )^{n-1\over2}  n !
  \ n^2
\end{equation}

    \begin{equation}C^{n+1}\build\sim_{n\rightarrow\infty}^{}
(-\ \delta_{\infty})^{n-1\over2} n !\, n^2
\end{equation}

 As far as the behaviour with respect to the external four momenta
 is concerned they follow
 the behaviour of the norm functions  \ref{1.8}, \ref{1.9} (i.e. the
corresponding structure of the Banach space).

 So despite the convergence of \emph{ ``$\Phi$-Iteration''},
(due to the alternating signs of the global
 terms)  the mapping:

 \begin{equation}{\cal M}:\, \ {\cal B}\stackrel{\cal M}
 \longrightarrow {\cal B}
\end{equation}

 defined by the above equations \ref{1.10}, \ref{1.11} is not contractive.
 As a matter of fact the $n^2$ dependence prevents the norms
from being conserved.

 This is the reason that motivated us to the definition of a new
mapping ${\cal M}^*$ (given
 by the following equations which is  \emph{contractive}:
            \begin{equation}H^{n+1'}(q,\Lambda) = {\delta_n^{'}(q,\Lambda)
 C^{n+1'}(q, \Lambda) \over 3\Lambda
 n (n-1)};\end{equation}
    with:\begin{equation}\delta_n^{'}(q,\Lambda)=\frac{3\Lambda n(n-1)}
 {(\gamma+\rho)+D_n(H)-\Lambda\alpha\Delta_F}\end{equation}
  and  \begin{equation}D_n(H)={|B^{n+1}| - |A^{n+1}| \over |H^{n+1}|}
\end{equation}

One can intuitevely understand the contractivity of the new
mapping ${\cal M}^*$  by looking at the
 behaviour with respect to $n$  of the function $D_n$
 (at fixed external momenta). Precisely,
    \begin{equation}D_n(H ) \build\sim_{n\rightarrow\infty}^{}{ n^2 n\ !
(\delta_{\infty}) ^{n-3\over2}\over  n\ !
(\delta_{\infty})^{n-1\over2}} \build\sim_{n\rightarrow
\infty}^{}{n^2\over\delta_{\infty}}.\end{equation}
Consequently one has:
\begin{equation}\delta_n^{'}(q,\Lambda)
\build\sim_{n\rightarrow\infty}^ {}{\delta_{\infty} n
(n-1)\over n^2}\build\sim_{n\rightarrow\infty}^{}
\delta_{\infty}
\end{equation}
By this last argument one can show not only the conservation of
the norms (in every dimension $r=0,1,2,3,4$)
 but also the contractivity  of  the ``new mapping'' ${\cal M}^*$
to a fixed point
 inside a characteristic subset $\Phi\subset {\cal B}$,
 under the following sufficient condition
imposed on the renormalized coupling constant:
\begin{equation}0\leq\Lambda\leq 0.01
\label{1.26}
\end{equation}
 In an equivalent way, this result means the existence and uniqueness
of a non trivial solution (even in four dimensions), of the
system. Under the condition \ref{1.26}
 this solution
lies in a neighbourhood of a precise point-sequence of the
appropriate subset $\Phi\subset {\cal B}$, the so called
\emph{fundamental sequence$\{H_0\}$}.

Consequently, the construction of this non perturbative solution
can be realized by iteration
 of the mapping ${\cal M}^*$ inside $\Phi\subset {\cal B}$
 starting from the corresponding to every dimension
\emph{fundamental sequence $\{H_0\}$}.
 So, this solution verifies automatically, the\emph{``alternating signs''}
 and
\emph{``splitting''} properties, at every value of the external
momenta,
 together with the
physical conditions imposed on $H^2$ and $H^4$-Green's
functions for the definition of the renormalization parameters.

 We note that the most important improvement of the method in $4$-dimensions as far as the corresponding properties of the solution are concerned \cite{MM5}
has been precisely the proof of the
 \emph{``alternating signs''} and \emph{``splitting''}
 properties, at every value of the external momenta,
 and not only at zero external momenta as we had originally
 established by the
 \emph{ ``$\Phi$-Iteration''} and the solution of the zero
dimensional problem.

\vspace{3mm}

                The reasons that motivated us for a study in smaller dimensions and
 not directly in four, were
 the absence of the difficulties due to the renormalization in two
 dimensions and the pure
 combinatorial character of the problem in zero dimensions.

 Another useful aspect  of the zero dimensional case is the fact that
 it provides a direct way to
test numerically the validity of the method.

\subsubsection{The last ``news'':}

  What  are the new developments at zero dimension:
  \begin{enumerate}
    \item The structure of the subset $\Phi \subset {\cal B}$
 characterized by the \emph{signs, ``splitting'' (factorization properties)} and bounds of the Green's
 functions $\left\{ H^{n+1} \right\}$ sequences in $\Phi$, 
 is reffined and more explictly described by a new subset
 $\Phi_0\subset \Phi$ 
 in terms of the
    "splitting" sequences upper $\left\{ \delta_{n_{max}}(\Lambda) \right\}$ and
    lower $\left\{ \delta_{n_{min}}(\Lambda) \right\}$ envelops.
 \item The non triviality of the subset $\Phi$ is established
    in terms of a new basic sequence $\left\{ H_{0} \right\}$ that we also use to prove the local contractivity.
    The stability of $\Phi$ is proved as a consequence of 
    the stability of the subset $\Phi_0$ which is directly obtained  recurrently thanks to the explicit definitions of the
    "splitting" sequences  $\left\{ \delta_{n_{max}}(\Lambda) \right\}$ and
  $\left\{ \delta_{n_{min}}(\Lambda) \right\}$.
  \item{}  The proof of contractivity is simpler in comparison with our previous
proofs also due to the fact that we introduce a new norm on the Banach space $\mathcal{B}$.

    \item Starting from the fundamental sequence $\left\{ H_{0} \right\}$ we define a
    new $\Phi^4_0$ iteration and explore numerically in Ref. \cite {MMST}
 the behavior of the Green's functions
    (essentially the $\delta$ functions), at sufficiently large $n$
    and reasonable order of this new iteration.

     The convergence is rapidly established for
    different values of $\Lambda$ and the sign and  splitting
    properties give consistent results with our
    theoretical conclusions
  \end{enumerate}

\section{The existence and uniqueness of the $\Phi^4_0$ solution}

\subsection{The vector space ${\cal B}$ and the $\Phi^4_0$\,
 equations of motion}

\begin{definition} {The space ${\cal B}$.}\ \label{def.2.1}

\emph{We introduce the vector space {${\cal B}$}
of the sequences $H=\{ H^{n+1}\} _{n =2k+1;  k\in\Bbb N}$
 by the following:}

\emph{The functions $H^{n+1}(\Lambda)$ belong to the space
 $C^{\infty}(\Bbb R^+ )$
 of continuously
differentiable numerical functions of the variable $\Lambda \in
\Bbb R^+$
 (which physically represents the coupling constant).}

 \emph{Moreover, there exists a universal (independent
 of n and of $\Lambda$)
 positive constant $K_0$, such that the following uniform
 bounds are verified:}

$$\forall  \ n=2k+1, k\in \Bbb N $$
\begin{equation}
|H^{n+1}(\Lambda)|\leq n\ !  (K_0)^n\qquad \forall\, \Lambda \in
\Bbb R^+
\end{equation}

\end{definition}

Let us now present the zero dimensional analog of the system of
equations of motion $(1.2.10... 14)$
 introduced in the previous chapter.

We suppose that the system of equations under consideration,
 concerns always
  the sequences of Euclidean connected
 and amputated with respect to the
free propagators Green's functions (the Schwinger functions). and
that these sequences  denoted by $H=\{ H^{n+1}\}_{n =2k+1\;,
k\in\Bbb N}$ belong to the above space {${\cal B}$}.

Taking into account the facts that in the present zero-dimensional
 case all the external
 four-momenta are set equal to zero, that the physical mass
 can be taken equal to $1$
 and that the
 renormalization parameters must be set equal to their trivial values namely:
\begin{equation}
 \alpha =0 = \rho;\, \,   \gamma=1,
\end{equation}
 one directly obtains the corresponding
 infinite system of {equations of
 motion
 for the sequence of the Schwinger functions} in the following form:

$\forall \ \Lambda\, \in\Bbb R^{+}$
  \begin{equation}  H^2 (\Lambda) =  - \Lambda H^4(\Lambda) \ \ +1
\end{equation}
 and for all $n \geq3$,
\begin{equation}
H^{n+1}(\Lambda) = \  \ A^{n+1}(\Lambda) +  B^{n+1}(\Lambda) +
C^{n+1}(\Lambda)
 \end{equation}

with:
\begin{equation}
A^{n+1}(\Lambda) =\,  - \Lambda H^{n+3}(\Lambda);
\label{1.1.6a}
\end{equation}
\begin{equation}
B^{n+1}(\Lambda) =\,  - 3\Lambda\sum_{\varpi_n(J)}{n \ !\over
j_{1}!j_{2}!} H^{j_{2}+2}(\Lambda) H^{j_{1}+1}(\Lambda);
\label{1.1.6}
\end{equation}
\begin{equation}
C^{n+1}(\Lambda) =\, - 6\Lambda\sum_{\varpi_n(I)} {n\ !\over
i_{1}!i_{2}!i_{3}!\ \sigma_{sym}(I)} \prod_{l=1,2,3}H^{i_{l}+1}
(\Lambda) \label{1.1.7}
\end{equation}

Here the notation $\varpi_n(J)$, means the set of different
partitions
 $(j_{1};j_{2})$ of $n$ such
that  $j_{1}$ is an odd integer and $j_{1}+j_{2}=n$.
 Respectively  $\varpi_n(I)$ is the set of
 triplets of odd numbers-different ordered partitions $(i_{1};i_{2};i_{3})$
 of $n$ with $ i_{1}\geq
i_{2}\geq i_{3}$.

The symmetry-integer $\sigma_{sym}(I)$ is defined by:
\begin{equation}
\sigma_{sym}(I) = \left\{\begin{matrix} \hfill 3!& if&
i_{1}=i_{2}=i_{3}\cr
                        \hfill  1& if&   i_{1}\neq i_{2}\neq i_{3}\neq i_{1}\cr
                     \hfill&2& otherwise  
                     \end{matrix}\right\}
\end{equation}
\vspace{5mm}
\subsection{The subset $\Phi \subset{\cal B}$}
\subsubsection{The splitting sequences}
\begin{definition} \ \

 We first introduce the class {${\cal D}$} of sequences
 $$\delta= \{\delta_{n}
 (\Lambda)\}
_{n=2k+1; k\in\Bbb N}\in {\cal B},$$
 such that they verify the bounds $(2.1.1)$ in the following simpler
 form:
\begin{equation}
|\delta_{n} (\Lambda)|\leq \,   K_0  ,
\end{equation}
$$   \forall\,  n=2k+1;\  k\in \Bbb N  $$
\end{definition}

\begin{definition} \textbf{{{splitting and signs in
$\Phi $}}}\label{def.2.3}\
\

  A sequence $H \in {\cal B}$ belongs to
the subset {$\Phi \subset{\cal B}$}
 if there exists an increasing associated sequence
 of positive and bounded functions on
 $\Bbb R^+$,
$$\delta= \{\delta_{n} (\Lambda)\}
_{n=2k+1; k\in\Bbb N}\in {\cal D},$$
 such that  the following
 ``splitting'' (or factorization) and sign
 properties are verified:$$\forall \Lambda \in \Bbb R^+$$ 

\begin{description}
\item[$\Phi.1$]
 \begin{equation}
  H^2 (\Lambda) =  1  +\Lambda \delta_1 (\Lambda)\,  \hbox{ with:}\,
  \,   \build\lim_{\Lambda \rightarrow 0}^{}
\delta_1 (\Lambda) = 0 \label{1.2.10}
\end{equation}

\item[$\Phi.2$]
 \begin{equation}   H^4 (\Lambda) =  -  \delta_3 (\Lambda)
\lbrack     H^2 (\Lambda)\rbrack^3,\,    \hbox{ with:}\,  \,
  \delta_3 (\Lambda)\leq 6\Lambda ,\   \build\lim_{\Lambda \rightarrow 0}^{}
{\delta_3 (\Lambda)\over \Lambda} = 6 \label{1.2.11}
\end{equation}

\item[$\Phi.3$]
\begin{equation}
\forall n\geq 5 \qquad
    H^{n+1} (\Lambda) = {\displaystyle  \delta_n (\Lambda)  C^{n+1}\over \displaystyle
 3\Lambda n(n-1)}
 ,  \  \hbox{ with :}\,  \  \build\lim_{\Lambda \rightarrow 0}^{}
 {\displaystyle\delta_n (\Lambda)\over
\displaystyle \Lambda} =  3n(n-1) \label{1.2.12}
\end{equation}

\item[$\Phi.4$]\ \

$\forall\, \, n=2k+1$\, with\,  $k\in\N,\, \exists$
  positive, (increasing with respect to $n$),  
 continuous functions of $\Lambda$,\, $$ \delta_{n,max}(\Lambda),\
 \delta_{n,min}(\Lambda),$$ 
 uniform limit and bound at infinity independent on $\Lambda $: $\hat \delta_{\infty} $,
 such that: 
 \begin{equation}
 \delta_{n,min}(\Lambda)< \ \delta_{n,max}(\Lambda)
\label{1.2.13}
\end{equation}
\begin{equation}
\delta_{n,min}(\Lambda)\leq \delta_{n}(\Lambda)
\leq\delta_{n,max}(\Lambda) \label{1.2.14}
\end{equation}
\begin{equation}
\build\lim_{n \rightarrow \infty}^{}\delta_{n,max} \leq
\hat \delta_{\infty} 
\label{1.2.15}
\end{equation}
\end{description}
\end{definition}

\begin{guess}\ \

\begin{description}
  \item[a.]
\emph{ We first remark that the above ``splitting'' or
factorization properties
    $\Phi.1,2,3$ are general formulae which simply define the function
 $\delta_{n} (\Lambda)$ in terms
 of the Green's function $H^{n+1}$ and the ``tree'' function $ C^{n+1}$.
 In other words, for every
  sequence $ H\subset {\cal B}$  the  corresponding splitting formulae
 can be formally
    written.
 The particular character of the subset $\Phi$ comes from
 the fact that the sequence of
 positive
 continuous functions $\delta= \{\delta_{n}
 (\Lambda)\}
_{n=2k+1; k\in\Bbb N}\in {\cal B}$  belongs to the class  ${\cal
D}$
 of uniformly bounded functions in ${\cal B}$
  and verifies the limit and asymptotic properties  $(\Phi.1, 2, 3, 4)$}

\item[b.] \emph{ The sequences $\{\delta_{n,max} (\Lambda)\}
_{n=2k+1; k\in\Bbb N}$ and $\{\delta_{n, min} (\Lambda)\}
_{n=2k+1; k\in\Bbb N}$ are not uniquely defined.  One can choose
many
 such sequences, and of course all of them yield an analogous structure
 of the subset $\Phi$}.
\emph{Nevertheless, we specify below such a couple of sequences
(cf. definition \ref{def.2.4})  
for the following  reasons:} 
 \begin{itemize}
\item{a)}\emph{  We shall use these specific splitting sequences: 
 $\{\delta_{n,max} (\Lambda)\}
_{n}$ and $\{\delta_{n, min} (\Lambda)\}
_{n}$ in order
to   construct the sequences: $$\{H^{n+1}_{max} \}
_{n=2k+1; k\in\Bbb N}\in\  {\cal B}\ \ \,  \hbox{and}\, \ 
 \{  H^{n+1}_{min}\}
_{n=2k+1; k\in\Bbb N}\, \in {\cal B}$$ and the ``fundamental sequence'' $\{H_0\} \in {\cal B}$
 (cf. definitions \ref{def.2.5} and  \ref{def.2.6} below).
 These sequences define an appropriate neighborhood, the subset} 
$\Phi_{0}\subset  \Phi$.
\item{b)}
\emph{We then show the stability of this subset  under the mapping 
${\cal M}^*$ and the local contractivity of
${\cal M}^*$ inside a ball  centered at $\{H_0\}$.}
\item{c)}
 \emph{A new $\Phi ^4_0$ iteration is defined, with starting point the 
sequence $\{H_0\}$ and in the reference
\cite{MMST},  we ``constructed numerically'' the solution of the 
dynamical system of equs. of the $\Phi^4_0$ model}.
\end{itemize} 
\end{description}
\end{guess}

\subsubsection{The ``neighborhood'' $\Phi_{0}\subset \Phi$}
\begin{definition}
\begin{equation}
    \delta_{3, max}(\Lambda) =6\Lambda;\qquad
 \delta_{3, min}(\Lambda) = {6\Lambda\over 1+9\Lambda}
\label{1.2.17}
 \end{equation}
\emph{and}\,  $\forall\, \,  n\geq 5 $
  \begin{equation}\delta_{n, max}(\Lambda)\, = {3\Lambda\ n(n-1)\over 1+ 3\Lambda\ n(n-1)d_0 }
 \end{equation}
 where   $d_0=0.01$
\begin{equation}\delta_{n, min}(\Lambda)\, = {3\Lambda\ n(n-1)\over 1+3\Lambda\
 n\ (n-1) }
\end{equation}
\label{def.2.4}
\end{definition}

\begin{definition} \ \ \label{def.2.5} 
We first define:
 \begin{equation}
  H^2_{max}(\Lambda )=(1+6\Lambda^2)^2;\
  \qquad   H^2_{min} =1
  \end{equation}
\begin{equation}
  H^4_{max}(\Lambda )=- 6\Lambda  [1+6\Lambda^2]^6;\ \
H^4_{min}
(\Lambda )= -{6\Lambda\over 1+9\Lambda}
\end{equation}
 Then, we define recurrently the following sequences, for every  $n=2k+1$ with  $ k\in\Bbb N, \  k\geq 2$,  by using  $\{\delta_{n, max}\} $\, and\,
 $\{\delta_{n, min}\}$
 introduced above: 
 \begin{description} 
 \item{(i.)}
\begin{equation}
  H^{n+1}_{max}= {\delta_{n,max} (\Lambda) C^{n+1}_{max}\over 3\Lambda\
n(n-1)}
\end{equation}
\begin{equation}
  H^{n+1}_{min}(\Lambda)= {\delta_{n,min}(\Lambda)  C^{n+1}_{min} \over
3\Lambda\ n(n-1)}
\end{equation}

Here  $C^{n+1}_{max},  C^{n+1}_{min}$, the so called ``tree terms'',  are defined by analogy to the definition \ref{1.1.7} of the introduction, precisely:
\begin{equation}
C^{n+1}_{max} =\, - 6\Lambda\sum_{\varpi_n(I)} {n\ !\over
i_{1}!i_{2}!i_{3}!\ \sigma_{sym}(I)} 
\prod_{l=1,2,3}H^{i_{l}+1}_{max}
\end{equation}
\begin{equation}
C^{n+1}_{min}=\, - 6\Lambda\sum_{\varpi_n(I)} {n\ !\over
i_{1}!i_{2}!i_{3}!\ \sigma_{sym}(I)} 
\prod_{l=1,2,3}H^{i_{l}+1}_{min}
\end{equation}

\item{(ii.)}\ In an analogous way we define: 
\begin{equation}
A^{n+1}_{max} =\,  - \Lambda H^{n+3}_{max};\qquad  A^{n+1}_{min} =\,  - \Lambda H^{n+3}_{min}.
\label{2.2.53}
\end{equation}
\item{(iii.)}
\begin{equation}
B^{n+1}_{max}(\Lambda) =\,  - 3\Lambda\sum_{\varpi_n(J)}{n \ !\over
j_{1}!j_{2}!} H^{j_{2}+2}_{max} H^{j_{1}+1}_{max};
\end{equation} 
\begin{equation}
B^{n+1}_{min}(\Lambda) =\,  - 3\Lambda\sum_{\varpi_n(J)}{n \ !\over
j_{1}!j_{2}!} H^{j_{2}+2}_{min} H^{j_{1}+1}_{min}.
\label{2.2.55}
\end{equation} 
\end{description}
\end{definition}

\vspace{5mm}


Now, we introduce the \textbf{``fundamental sequence''.}

\begin{definition}\label{def.2.6}
 \begin{equation}
    H^2_0 (\Lambda) =  1-\Lambda H^4 _{min}
    \ \ \
    \end{equation}
    \begin{equation}
   H^{4}_{0} (\Lambda) = - \delta_{3, 0}(\Lambda)[H^{2}_{0}]^3\quad
\hbox{with}\quad \delta_{3, 0} (\Lambda)= \displaystyle{ \frac{6\Lambda}{1+9\Lambda H^{2}_{min}
 - \displaystyle{\frac{\Lambda
|H^6_{min}|}{ |H^4_{max}|} }}}
\label{1.2.52}
\end{equation}
and for every $n\geq 5$
    \begin{equation}
H^{n+1}_{0}(\Lambda) = {\delta_{n, 0}(\Lambda)
 C^{n+1 }_{0}( \Lambda) \over 3\Lambda
 n (n-1)};
\label{1.2.53}
\end{equation}

      with:
\begin{equation}C^{n+1}_0(\Lambda) = - 6\Lambda\sum_{\varpi_n(I)}
 {n\ !\over i_{1}!i_{2}!i_{3}!\
 \sigma_{sym}(I)}
\prod_{l=1,2,3}H^{i_{l}+1}_0 (\Lambda);
\end{equation}

\begin{equation}
\delta_{n,0}(\Lambda)=\frac{3\Lambda n(n-1)}
 {1+D_{(n,min)}(H)}
\label{1.2.54}
\end{equation}
and
 \begin{equation}
D_{(n,min)}(H)=\displaystyle{\frac{|B^{n+1}_{min}|}{|H^{n+1}_{min}|}-\frac{|A^{n+1}_{max}|}{|H^{n+1}_{max}|}}
 \label{1.2.55}
\end{equation}
Here the terms $B^{n+1}_{min}$ and  $A^{n+1}_{max}$  are defined by the definitions \ref{2.2.55} and \ref{2.2.53} respectively.
\end{definition}

\begin{definition}\ \textbf{The subset} $\Phi _0$\label{def.2.7}

Taking into account the sequences of definitions \ref{def.2.4} and \ref {def.2.5} we introduce
 the following subset $\Phi_0\subset \Phi$:
\begin{equation}
\Phi _0=\left\{ H\in\Phi : \vert H^{n+1}_{min}\vert\ \leq
\vert H^{n+1}\vert\leq
\vert H^{n+1}_{max}\vert , \quad  \forall n=2k+1,\ k\in\ \N\right  \}
\label{3.60}
\end{equation}
\end{definition}

\vspace{4mm}

\begin{proposition}{ {(the non triviality  of $\Phi_0$  and $\Phi$)}}

The set of sequences {$\Phi_0$} given by the
definition  \ref{def.2.7} is a nontrivial subset of the space
 ${\cal B}$. 
\label{prop.2.1}
\end{proposition}

\vspace{3mm} 

{\bf Proof of Proposition \ref{prop.2.1}}

 From the  previous definitions of the sequences $\{\delta_{n,max} \} 
 \,\ \ 
 \{\delta_{n, min} \}
$  (cf. definition \ \ref{def.2.4}) and of  the sequences $\{H^{n+1}_{max} \},
 \, \ 
 \{  H^{n+1}_{min}\}$,   $\{H_0\}$
 (cf. definitions \ref{def.2.5} and  \ref{def.2.6}), and the  obvious strict inequality  $$\delta_{n,max} (\Lambda)
> \ \delta_{n, min} (\Lambda)\ \  	 
\forall  \ n=2k+1; k\in\Bbb N^* $$ one directly ensures that:
\ \begin{itemize}
\item{a)} The previously defined fundamental   sequence $H_0$
satisfies the properties  $(\Phi.1, 2, 3, 4)$ of definition \ref{def.2.3} and  definition \ref{def.2.7} for every $\Lambda$
 in the interval $]0; 0.05]$
\item{b)} The  set $\Phi_0$ is a non trivial subset of the set $\Phi$, which in turn is a non trivial subset of ${\cal B}$.\   $\blacksquare$

\end{itemize}

\vspace{3mm}
Let us now present some definitions and auxiliary results, that one could easily obtain from the previous  definitions.
\begin{proposition}\label{prop.2.2}
  $\forall n\geq 5$ the tree terms $C^{n+1}_{min}$ and $C^{n+1}_{max}$ verify:
\begin{equation}\begin{array}{l}\vert C^{n+1}_{min}\vert\geq \ \vert\bar C^{n+1}_{min}\vert\\
\ \ \\
\vert C^{n+1}_{max}\vert\leq \ \vert\bar C^{n+1}_{max}\vert
\end{array}$$
where: 
$$\begin{array}{l}
\vert\bar C^{n+1}_{min}\vert=\mathcal{T}_{n} \vert C^{n+1}_{(n/3,n/3,n/3)(min)}\vert\\
\ \ \\
\vert\bar C^{n+1}_{max}\vert=\mathcal{T}_{n} \vert C^{n+1}_{(n-2, 1, 1)(max)}\vert
\end{array}\end{equation}
Where we have to take into account  the result
of ref. \cite[c]{MM1} about the number $\mathcal{T}_{n}$
 of different partitions
of the set $\varpi_{n}(I)$ in the sum
$\sum\limits_{\varpi_{n}(I)}$
precisely:\\
for every $n\geq 9$~:
\begin{equation}
    \mathcal{T}_{n}=\frac{(n-3)^{2}}{48}+\frac{(n-3)}{3}+1
\label{V.21}
\end{equation}

\end{proposition}

\subsubsection{The ``sweeping factors''}
\begin{definition}
 \label{def.2.4 }\emph{ For every $H\in \Phi_0$, we
introduce the sequence of functions,}
$$\{Y_{n}(H)\}
_{n=2k+1; k\in\Bbb N^*}\in {\cal D}\subset {\cal B}$$ \emph{called
the} ``sweeping factors'' \emph{and defined as follows:}
\begin{equation}
Y_3=\frac{1}{6} ; \quad Y_5= \frac{1}{20} 
 \end {equation}
\emph{and for every} $n\geq7$
\begin{equation}
Y_{n}(H)=\frac{-C^{n+1}}{3\Lambda n^2 (n-1)^2 H^{n-1}[H^2]^2}
\label{1.2.33}
\end{equation}
\label{def.1.5}
\end{definition}

\vspace{5mm} 
The limits at $\Lambda\rightarrow\ 0$ and positivity  of
the coupling constant and that of the
 splitting sequences
implies directly (in view of the previous definitions) the limits, the
\emph{``good sign''} properties $(-1)^{\frac{(n-1)}{2}}$ of the
 Schwinger functions,
the \emph{complete ``splitting''} (or \emph{factorization})
properties, together with the combinatorial asymptotic behaviour
of the
 $H^{n+1}$ functions. We present these results without proof
 by the following statement.

\subsubsection{``Alternating  signs'' and ``complete `splitting`'' } 
\begin{proposition}\  \
Let $H\in\, \Phi_0 $, then for every $\Lambda \in]0; 0.035]$
\begin{itemize}
\item{}
\begin{equation}
\lim_{\Lambda \rightarrow0}H^2 (\Lambda)=1,\qquad \lim_{\Lambda
\rightarrow 0} \frac{H^4 (\Lambda)}{\Lambda}=-6 \label{1.2.36}
\end{equation}
and
\begin {equation}
\forall\, n\geq 5,\quad \lim_{\Lambda \rightarrow 0} \frac{H^{n+1}
(\Lambda)}{(-\Lambda)^{(n-1)/2}}=n!c_n \label{1.2.37}
\end{equation}
where the positive constants $c_n$ are recurrently defined by:

\begin{equation}c_n=6 \sum_{\varpi_n(I)}\frac{\prod_{j=1, 2,
3}c_{i_j}}{\sigma_{sym}(I)} \qquad \hbox{with}\quad c_3 =1=c_1.
\end{equation}

\vspace{8mm}

\item{} for
every $n\geq 3$ we have:
\begin {description}
\item{a)}\begin{equation}
 H^{n+1}(\Lambda)= - n(n-1)\delta_n (\Lambda)\, Y_n (H) H^{n-1}[H^2]^2
\label{2.2.44}
\end{equation}
\item{b)}
\begin{equation}
 H^{n+1}(\Lambda)= n! (-1)^{\frac{(n-1)}{2}}\,
 \displaystyle{[H^2]^n}\prod_{m=3}^{n}
 Y_m (H)\delta_m (\Lambda)
\label{2.2.48}
\end{equation}
\end{description}
\end{itemize}
\end{proposition}

\vspace{3mm}

\section{The new mapping ${\cal M^*}$ and the stability of $\Phi_0$}

We now define the new mapping ${\cal M^*}$ and then prove the
stability of the subset
 $\Phi_0$ (and consequently of $\Phi$) under the action of this mapping.

\vspace{3mm}


\begin{definition}\, {The new mapping ${\cal M^*}$}\ 
\label{def.3.1}.

Let $H\in \ \Phi_0$
We define the following application
 ${\cal M^*}:\, \Phi_0 \stackrel{\cal M^*}
\longrightarrow
 {\cal B}$ by:

\begin{equation}
H^{2'} (\Lambda) = 1-\Lambda H^4 (\Lambda)
\label{1.2.51}
\end{equation}

 \begin{equation}
H^{4'} (\Lambda) = - \delta_3'(\Lambda)[H^{2'}]^3\quad
\hbox{with}\quad \delta_3^{'}(\Lambda)= \frac{6\Lambda}{1+\ D_3}
\end{equation}
\begin{equation}\mbox{and}\quad D_3=6\Lambda
H^2\left(3/2
 - \frac{|H^6|}{6|H^4||H^2|}\right).
\end{equation}
Moreover:
\begin{equation}
\forall\  n\geq 5 \qquad
    H^{n+1'} (\Lambda) = \displaystyle{ \delta_n^{'}(\Lambda)  C^{n+1'}\over \displaystyle
 3\Lambda n(n-1)}
  \label{1.2.12}
\end{equation}
Here: \begin{equation}\delta_n^{'}(\Lambda)=\frac{3\Lambda n(n-1)}
 {1+D_n(H)}
\label{1.2.54}
\end{equation}
 with:
 \begin{equation}
D_n(H)={|B^{n+1}| - |A^{n+1}| \over |H^{n+1}|} \label{1.2.55}
\end{equation}
\end{definition}

\begin{guess}\
 \emph{Notice that in the above definition  the functional $D_n(H)$
 is obtained in terms
of the absolute values of the Green's functions, thanks to
 the hypothesis $H\in\Phi_0\subset\ \Phi$ which implies the good sign properties of the Green's functions.}
 \end{guess}
\vspace{3mm}

Taking into account the infinite system of equations of motion of
the Green's functions (cf. 2.1.3...7), the structure of the subset
$\Phi$ (cf.def.2.3)
 and the above definition \ref{def.3.1}, one easily verifies the equivalence of
the mappings ${\cal M}$ and ${\cal M^*}$.

\vspace{3mm}


\begin{axiom}\label{Th.3.1}
{The stability of the subset $\Phi_{0}\subset{\Phi}$}\ 

Let $H\in \Phi_{0}$ then: $${\cal M^*}(H)\subset \Phi_0$$
  under the condition:
$$0<\Lambda\leq 0.05$$
\end{axiom}

 The proof  of the stability of $\Phi_0$ under the mapping ${\cal M^*}$  is given in Appendix I of section \ref{Ap.1} by using the following four auxiliary propositions also established in this Appendix.
 
\vspace{3mm}

\begin{proposition}{Signs and bounds for  $n=1,\  3,\ 5$ }\label{prop.3.1}\ \

Let ${H\in \Phi_0}$ then \ \ $\forall \Lambda\leq 0.05$:

\begin{itemize}
\item[i)] \begin{equation}
H^{2'}>0;\ \quad H^{4'}<0, \quad  H^{6'}>0.
\end{equation}
\item[ii)] \begin{equation} 
\begin{array}{l} 
H^{2}_{min}<H^{2'}<H^{2}_{max}\\
\ \ \\
|H^{4}_{min}|< \  |H^{4'}| <|H^{4}_{max}|\\
\ \ \\
H^{6}_{min}<H^{6'}<H^{6}_{max}
\end{array}
\end{equation} 
\end{itemize}
\end{proposition}

\begin{proposition}\label{prop.3.2}\

Let ${H\in \Phi_0}$ then \ \ $\forall \Lambda\leq 0.05$:

The functional $A^{n+1}_{}(H)$ given by definition
 \ref{def.2.1} verifies the following properties:
\begin{itemize}
\item[i.] \textbf{the ``good sign'' property:}
$$\forall \ n=2k+1\  (k\geq 1)\ \  A^{n+1}=(-1)^{\frac{n-1}{2}}|A^{n+1}| $$
\item[ii.] $\forall \ n=2k+1\  (k\geq 3)$, the sequence:
$$\displaystyle{\frac{|A^{n+1}_{max}|}{ n(n-1|H^{n+1}_{max}|}}$$
decreases with increasing $n$.
\end{itemize}  
\end{proposition}

\vspace{1mm}

\begin{proposition} \label{prop.3.3}\

Let ${H\in \Phi_0}$ then \ \ $\forall \Lambda\leq 0.05$:

The functional $B^{n+1}(H)$  given by definition
 \ref{def.2.1} verifies the following properties:
\begin{itemize}
\item[i.] \textbf{the ``opposite sign'' property:}
$$\forall \ n=2k+1, \  (k\geq 1)\ \  B^{n+1} =(-1)^{\frac{n-3}{2}}|B^{n+1} |$$ 
\item[ii.] $\forall \ n=2k+1\  (k\geq 3)$, the sequence:
$$\displaystyle{\frac{|B^{n+1}_{min}|}{n(n-1)|H^{n+1}_{min}|}}$$
increases with increasing $n$.
\end{itemize} 
\end{proposition}

\begin{proposition}\label{prop.3.4} \

Let ${H\in \Phi_0}$ then \ \ :

$\forall n\geq 7$
\begin{description}
\item{i)}  $$\displaystyle{\lim_{\Lambda\rightarrow 0}D_n(\Lambda)}=0$$
\item{ii)} $\forall \Lambda\leq 0.05$ the sequence:
$$\displaystyle{\frac{D_{(n,min)}(H)}{3\Lambda n(n-1) }}$$
increases with increasing $n$.
\item{iii)} There exists a finite  positive real number $1>  d_n >\ d_0$  \ such that:
 $$\displaystyle{\frac{1}{2}\geq \frac{ D_n(H) }{3\Lambda n(n-1) }}\geq   d_n\quad \mbox{and} \ \ \displaystyle{\lim_{n \to \infty}d_n=d_0=\frac{1}{\delta_\infty}}$$
\emph{Reminder}:
 \begin{equation}
D_n(H)={|B^{n+1}| - |A^{n+1}| \over |H^{n+1}|} 
\end{equation}
\end{description}
\end{proposition}
\vfill\eject
\vspace{5mm}
\section{Construction of the  $\Phi^4_0$\, unique
 and non trivial solution}

\subsection{The Banach space}
In this third part of the study  of the $\Phi^4_0$ problem we
present the construction of the unique non trivial
 solution inside
the subset $\Phi$, by the following three steps:
\begin {description}
\item [i.] We introduce a precise norm ${\cal N}$ inside ${\cal
B}$ and show the closedeness and completeness of $\Phi_0$
 in the induced norm.
\item [ii.] We  prove the contractivity of ${\cal M} ^*$
 inside a neighborhood $\Phi_0$
 and consequently the existence and uniqueness of a fixed
 point of the initial mapping
${\cal M}$ inside this particular subset of ${\cal B}$. \item
[iii.] For the construction of the solution
 we realized (cf.\cite{MMST}) an iteration
 of the mapping ${\cal M} ^*$, starting from the
 particular sequences $\{\delta_{n(max)}\}$ and $\{\delta_{n(min)}\}$, which define the  neighborhood of the
 ``fundamental sequence'' $H_0$ introduced in the
 previous section.
\end{description}

\vspace{3mm}

 \begin{definition}\label{def.4.1}
 We introduce the following mapping ${\cal N}$ from ${\cal B}$
to $\Bbb R^+$:

\begin{equation}\begin{array}{l}
{\cal N}:\ {\cal B}\rightarrow \Bbb R^+\  \\
\ \ \ \ H\mapsto\|H\|\ \  \hbox{with:}\ \  \|H\| =  \displaystyle
{\sup_{\Lambda;n }}\displaystyle{\frac{|H^{n+1}|}{M_n}  
 } 
\end{array}
\label{3.1.94}
\end{equation}
Here $\forall\ \  \Lambda\in\Bbb R^+$:
\begin{equation}
M_1 (\Lambda)=H^2_{max}=(1+6\Lambda^2)^2;\quad M_3 (\Lambda)=\,
 \delta_{3,max}(M_1)^{3};\  \
\label{3.1.95}
\end{equation}
and for every $n\geq 5$

\begin{equation}\begin{array}{l}
M_n (\Lambda)=\ n(n-1) \delta_{n,max} M_{n-2}(M_1)^2\ \ \ \ \ \\
\ \ \ \ \\
\end{array}
\label{3.1.96}
\end{equation}
\label{def.4.1}
\end{definition}

\begin{guess}\ \
\begin{enumerate}

\item{}\emph{We note  that the above function ${\cal N}$
defines
 a finite
norm inside
 a non empty subspace ${\cal B}_c$ of ${\cal B}$.
 This subspace ${\cal B}_c$
evidently contains $\Phi$ (and $\Phi_0$) and it is a Banach space with
 respect to the ${\cal N}$- topology}.

\item{} \emph{Using the above definition \ref{def.4.1} of the norm
in ${\cal B}$, we introduce a ball-neighborhood $S_{\rho}(H_0)\subset \Phi_0$  of the
{fundamental sequence} and show by theorem \ref{Th.4.1}  which
follows, the local contractivity of ${\cal M}^* $
inside it.} 
\end{enumerate}
\end{guess}

\subsection{The local contractivity}

\begin{definition}
\begin{equation}
\begin{array}{l}
S_{\rho}=\{H\in\Phi_0: \Vert H-H_{0}\Vert \leq \rho \}\\
\ \ \\
\mbox{with} \ \  \rho=\displaystyle{\sup_{n}\frac{\{\ \delta_{n,max}-\delta_{n,min}\}}{\delta_{n,max}}} 
\end{array}
\end{equation}
\label{def.S}
\end{definition}
 In view of the definition \ref{def.2.4} of the sequences  $\{\ \delta_{n,max}\}\ \ \mbox{and}\ \ \  \{\delta_{n,min}\}$, we directly verify the following:
\begin{proposition}
\begin{equation}
\rho=\displaystyle{\lim_{n\to \infty}\frac{\{\ \delta_{n,max}-\delta_{n,min}\}}{\delta_{n,max}}} =1- d_0
\end{equation}
\end{proposition}

\vspace{3mm}
In Appendix  II we give in detail  the proof of the following:
\begin{axiom}  {\textbf{ { The local
contractivity} of the mapping
 ${\cal M}^* $}\ \ 
in $S_{\rho}(H_0)\ \subset \Phi_0$} \label{Th.4.1} \ 

\begin{description}
\item[i.] The subset (ball) $S_{\rho}(H_0)\ \subset \Phi_0$ is a complete metric
 subspace of ${\cal B}_c$
in the induced topology.
  \item[ii.] There exists a finite positive
constant
 $\Lambda^*(\approx 0.45)$
 such that the mapping ${\cal M}^* $ is locally contractive inside  $S_{\rho}(H_0)\ \subset
\Phi_0$.
\item[iii.] When
 $\Lambda\in]0, \Lambda^*]$  the unique non trivial solution of the $\Phi_0^4$
 equations of motion lies in the neihbourhood $S_{\rho}(H_0)\ \subset
\Phi_0$  of the fundamental sequence
$H_0$.

\end{description}
\end{axiom}

\vspace{5mm}
\textbf{Acknowledgments}

The author is grateful to J.Bros for his constant encouragement during the various phases of this work and for a critical reading of the manuscript.

\vfill\eject

\vfill\eject
\section{APPENDIX I}\label{Ap.1}
\subsection{Proof of propositions \ref{prop.3.1}, \ref{prop.3.2}, \ref{prop.3.3}, \ref{prop.3.4}}
 \begin{enumerate}
\item{\textbf{Proof of proposition \ref{prop.3.1}}}\
\begin{itemize}
\item{n=1} 
 \begin{equation}
 \begin{array}{l}
 \mbox{From \ref{def.2.5}, \ref{def.3.1} and     the hypothesis  we have}:\\
H^{2'}= 1-\Lambda H^{4} , 
 \  H^{4}=( -1)|H^{4}|\ \ \mbox{and} \  \ |H^{4}|\leq |H^{4}_{max}|\\
 \ \ \\
\Rightarrow\ \ H^{2'}>0\ \ \mbox{and}\ \ \   |H^{2'}|\leq  1+6\Lambda^2(1+6\Lambda^2)^6).\\ 
 \mbox{Then we verify that for every} \Lambda\leq 0.05\\
1+6\Lambda^2(1+6\Lambda^2)^6 \leq (1+6\Lambda^2)^2\\
\Leftrightarrow\ \  |H^{2'}|\leq  |H^{2}_{max}|\ \ \qquad \blacksquare\\Ê
\mbox{Respectively, from definition \ref{def.2.5}}:\ \ \ H^4_{min}(\Lambda )= -\displaystyle{{6\Lambda\over 1+9\Lambda}}\\
\ \ \Rightarrow |H^{2'}|\geq 1+\displaystyle{{6\Lambda^2\over 1+9\Lambda(1+6\Lambda^2)}}>1=  H^{2}_{min} \ \ \blacksquare\\
  \end{array}
\end{equation}
    
\item{n=3}\ \ 
 In an analogous way:
\begin{equation}
    \begin{array}{l} 
\mbox{from definitions \ref{def.2.5}, \ref{def.3.1} and     the hypothesis  we have:}
\ \ \\
H^{4'} (\Lambda) = - \delta_3'(\Lambda)[H^{2'}]^3\\
\hbox{with}\quad \delta_3^{'}(\Lambda)= \displaystyle{\frac{6\Lambda}{1+\ D_3}}\quad
\mbox{and}\quad D_3=6\Lambda
H^2(3/2
 - \displaystyle{\frac{|H^6|}{6|H^4||H^2|}}.\\
 \mbox{But,}\ \ \\
 \mbox{by the hypothesis}  \quad \displaystyle{\frac{|H^6|}{|H^4||H^2|}}\leq \displaystyle{\frac{\delta_{5max}|H^4||H^2|}{|H^4|}}\leq\displaystyle{\frac{60\Lambda |H^2_{max}|}{1+60\Lambda d_0}}\\
 \  \mbox{with}\ \  d_0=0.01 \ \\
\mbox{For the bounds and positivity of\ $\delta_3^{'}$
  it is sufficient to ensure that}\ \  D_3>0 \\
\mbox{or the stronger condition:} \\
 3- 0,6 \Lambda - 20 \Lambda (1+6\Lambda^2)^2>0
 \\
 \mbox{The latter is verified for}\ \   \Lambda\leq 0.05 \\
\mbox{Moreover trivially:}\quad 1+9\Lambda > D_3\\
\ \ \\
 \ \Rightarrow \ \quad  \delta_{3min} <  \delta_3^{'}<\ 6\Lambda =  \delta_{3,max}. \\
\mbox{And, by the previous results for the two-point function} \ \ (n=1)\\
 \mbox{we finally obtain that:}\ \ H^{4'}<0,\ \quad\ |H^{4}_{min}|< \  |H^{4'}| <|H^{4}_{max}|\qquad\qquad \blacksquare
 \end{array}
\end{equation}

\item{n=5}
 
  In an analogous way:
\begin{equation}
    \begin{array}{l} 
\mbox{from definitions \ref{def.2.5}, \ref{def.3.1} and     the hypothesis  we have:}
\ \ \\
H^{6'} (\Lambda) = = \displaystyle{ \delta_5^{'}(\Lambda)  C^{6'}\over \displaystyle
60\Lambda  }
\\
\hbox{with}\quad \delta_5^{'}(\Lambda)= \displaystyle{\frac{60\Lambda}{1+\ D_5}}\quad
\mbox{and}\quad D_5=\displaystyle{\frac{|B^6|}{|H^6|}-\frac{| A^6|}{|H^6|}}
\end{array} 
\end{equation}
We then show that:
\begin{equation}\begin{array}{l} 
\mbox{for fixed   $d_0=0.01$ and}\ \ \  
\forall  \ \Lambda \leq 0.05\\ 
\exists \ \  \mbox{a finite positive constant} \ d_5>d_0 \\
 \mbox{such that}   \ D_5> 60\Lambda d_5. 
 \end{array}
 \label{5.86}
 \end{equation}
In order that  the previous condition \ref{5.86} be satisfied, and in view of  the hypothesis $H\in  \Phi_0$ (application of bounds and signs in the expression of $D_5$) it is sufficient to verify the following stronger condition:
\begin{equation}\begin{array}{l}
(6+45\Lambda+60\Lambda d_0)(1+3\Lambda 42 d_0)\\
- 2\Lambda 42  H^2_{(max)}(1+9\Lambda)- 20 d_5(1+9\Lambda)(1+3\Lambda 42 d_0)> 0
\end{array}
\end{equation}
For fixed $d_0=0.01 $ and $\Lambda\sim 0.05$
the latter allows us to require the following stronger condition:
 \begin{equation}
 2.4>20d_5 \times 2.0197\ \ \ \mbox{which yields:}\ \ \ d_5\sim0,059>d_0\ \ \blacksquare
 \end{equation}
 Taking into account the latter together with the earlier results
 for $n=1$, and $n=3$ we finally obtain:
 $$H^{6'}>0,\ \quad\ |H^{6}_{min}|< \  |H^{6'}| <|H^{6}_{max}|\ \ \ \ \blacksquare$$  
 \end{itemize}
 
 \item{\textbf{Proof of proposition \ref{prop.3.2}}}\
 \begin{itemize}
\item[i.] 
 The sign property is directly obtained by the hypothesis $H\in\ \Phi_0$
  \begin{equation}
  \begin{array}{l}
  A^{n+1}= - \Lambda H^{n+3}  = -\Lambda (-1)^{\frac{n+1}{2}}|H^{n+3}|=\\
  =\Lambda (-1)^{ 2} (-1)^{\frac{n-1}{2}}|H^{n+3}| =(-1)^{\frac{n-1}{2}}|A^{n+1}| \ \ \blacksquare
  \end{array}
  \end{equation}
  
  \item[ii.] In order to ensure the decrease property we show that:
  \begin{equation}
  \displaystyle{\frac{|A^{n+1}_{max}|}{n(n-1)|H^{n+1}_{max}|}}<\ \displaystyle{\frac{|A^{n-1}_{max}|}{(n-2)(n-3)|H^{n-1}_{max}|}}
  \end{equation}
or in an equivalent way, by  taking into account the definitions of the splitting sequences and Green's functions in $\Phi_0$ \ref{def.2.4}, \ref{def.2.5}, as well as proposition \ref{prop.2.2}
(upper bounds of tree terms $C^{n+1}$ we require:
\begin{equation}
  \displaystyle{\frac{\delta_{n+2,max}{\cal T}_{n+2}}{n(n-1)}}<\ \displaystyle{\frac{\delta_{n,max}{\cal T}_{n}}{(n-2)(n-3)}}
  \label{5.91} 
   \end{equation}
or equivalently, by inserting the corresponding expression of the number ${\cal T}_{n}$ of different partitions in $C^{n+1}$ (cf.\ref{V.21}):
\begin{equation}
\begin{array}{l}
  \displaystyle{\frac{(n+1)(n+2)}{1+3\Lambda(n+1)(n+2)d_0}(1+\frac{15}{n}+\frac{48}{n(n-1)})}<\\
   \displaystyle{\frac{n(n-1)}{1+3\Lambda n(n-1)d_0}(1+\frac{15}{(n-2)}+\frac{48}{(n-2)(n-3)})}
   \end{array}
 \label{5.92}  
  \end{equation} 
  \vspace{2mm}
  \begin{figure}[h]
\begin{center}
\hspace*{-5mm}
 \includegraphics[width=12cm]{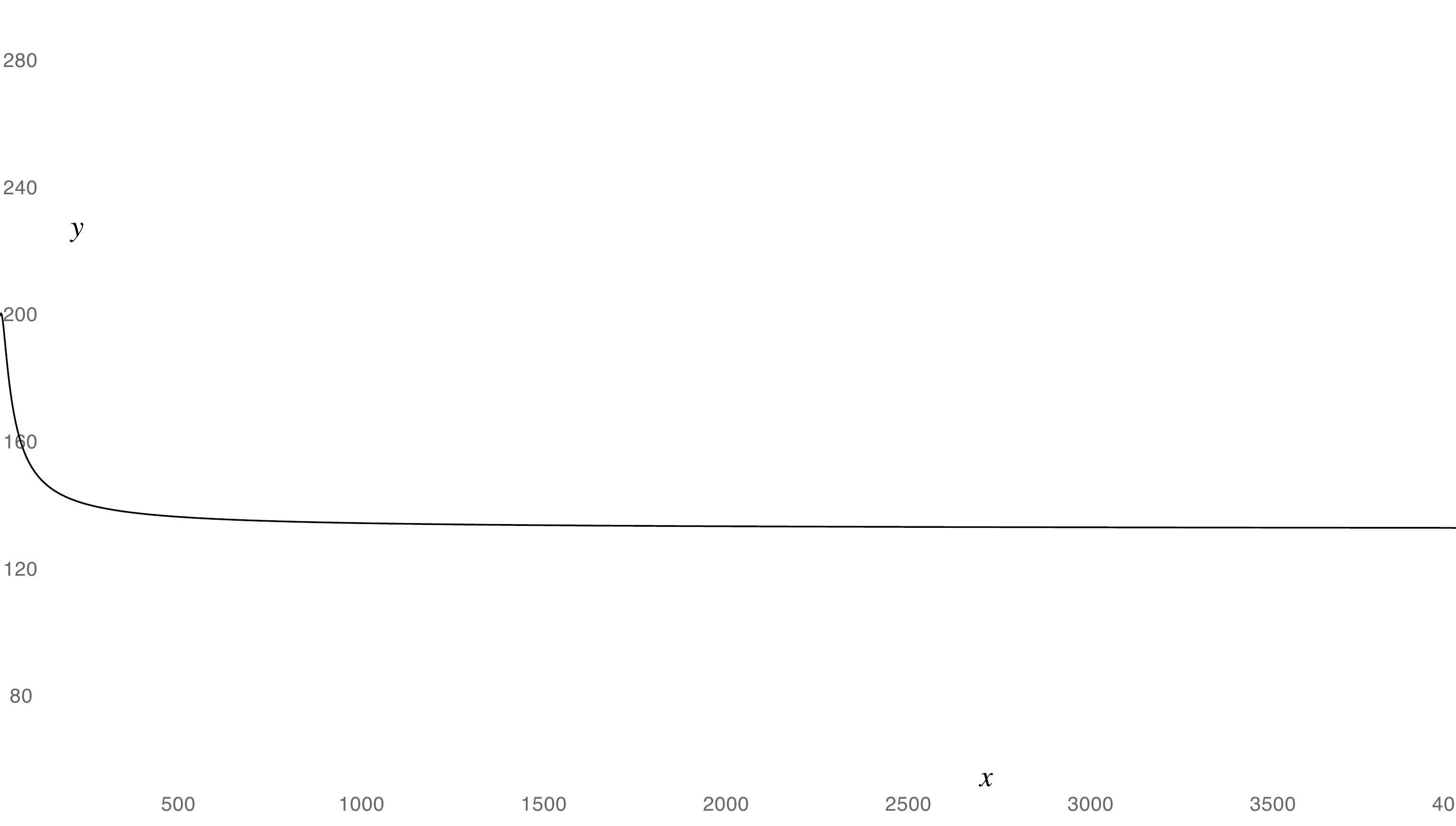}
\end{center} 
 \caption{\small\textrm{For the values of $n$ ($=x$ continuous) in the interval $]7, 4001] $ the function $f_{L}(x)$ decreases continuously up to the limit value of $(0,0075)^{-1} \sim 133.33\dots$}}
\label{fig.1.1}
\end{figure}

When we fixe $d_0=\Lambda=0,05$  we verify (after some elementary but long manipulations) that the dominant contribution ({\it{i.e.}} $\sim n^2$) of the function : 
\begin{equation}
\begin{array}{l}
 f(n)=f_{L}(n) - f_{R}(n)\\
 \mbox{where:}\\
 f_{L}(n)= \displaystyle{\frac{(n+1)(n+2)}{1+3\Lambda(n+1)(n+2)d_0}(1+\frac{15}{n}+\frac{48}{n(n-1)})} \ \ 
 \ \mbox{and}\\
f_{R}(n)=\displaystyle{\frac{n(n-1)}{1+3\Lambda n(n-1)d_0}(1+\frac{15}{(n-2)}+\frac{48}{(n-2)(n-3)})}\\
 \mbox{is strictly negative}\\
   \end{array}
  \end{equation} 
  More precisely we find:
  \begin{equation}
  \begin{array}{l}
 f(n)= 2(2n+1)+ \displaystyle{\frac{30 n [1-2/n-2/n^2]}{n-2}-\frac{192n [2-2/n-3/n^2]]}{(n-1)(n-2)(n-3)}}- \\
 -0.0075 \displaystyle{\frac{30n^2(n+1)(n+2) }{(n-2)(n-3)}}[1+72/30n-198/30n^2] 
 \end{array}
  \end{equation}

 Of course, an analogous result is also  obtained  if we consider 
  that $n=x$ (as a continuous variable). The derivative of the left hand side of the inequality  \ref{5.92} is negative when we fixe $d_0\sim\Lambda\leq 0.05$ and for $x >7$.
  
 In figure \ref{fig.1.1} we graphically illustrate the behaviour of the function: 
 $$f_{L}(n)= \displaystyle{\frac{(n+1)(n+2)}{1+3\Lambda(n+1)(n+2)d_0}(1+\frac{15}{n}+\frac{48}{n(n-1)})}$$
 We notice that when  $n=x$ varies continuously from $x=7$  up to $x= 4001$, $f_{L}(n)$ decreases continuously up to the limit value of $(0.15\times 0.05=0,0075)^{-1} \sim 133.33\dots$. 
This ends the proof of proposition \ref{prop.3.2}. \ \ \ $\blacksquare$
   \end{itemize}

 \item{\textbf{Proof of proposition \ref{prop.3.3}}}
  \begin{itemize}
\item[i.] 
 As previously, the sign property is directly obtained by the hypothesis $H\in\ \Phi_0$:
  \begin{equation}
  \begin{array}{l}
    B^{n+1}(\Lambda) =\,  - 3\Lambda\displaystyle{\sum_{\varpi_n(J)}{n \ !\over
j_{1}!j_{2}!} H^{j_{2}+2}(\Lambda) H^{j_{1}+1}(\Lambda)}=\\
  =- 3\Lambda \displaystyle{\sum_{\varpi_n(J)}{n \ !\over
j_{1}!j_{2}!} (-1)^{\frac{j_2}{2}}| H^{j_{2}+2}|(-1)^{\frac{j_1-1}{2}} |H^{j_{1}+1}|}\\
 =(-1)^{\frac{n-3}{2}}|B^{n+1}| \ \ \blacksquare
  \end{array}
  \end{equation}
  \item[ii.] In order to ensure the increase property we show that:
  \begin{equation}
  \displaystyle{\frac{|B^{n+1}_{min}|}{n(n-1)|H^{n+1}_{min}|}}>\ \displaystyle{\frac{|B^{n-1}_{min}|}{(n-2)(n-3)|H^{n-1}_{min}|}}
  \label{5.96}
  \end{equation}
  The left hand side of \ref{5.96} can be expressed in terms of the smaller contribution $B^{n+1}_{(j_1, j_2), min}$ (with $j_{1}=\frac{n-1}{2}, \ j_2= \frac{n+1}{2}$)  times the number of different partitions $\varpi_n(J)$ in the sum of $B^{n+1}_{min}$ $\it{i.e.} \frac{n-1}{2}$; in other words: 
  $$\frac{n-1}{2}B^{n+1}_{(j_{1}=\frac{n-1}{2},  j_2= \frac{n+1}{2})}$$
  Reminder: $$B^{n+1}_{(j_{1}=\frac{n-1}{2},  j_2= \frac{n+1}{2})}=\displaystyle{{n ! H^{(j_1+1)}H^{(j_2+2)}\over
((n-1)/2)!((n+1)/2)!}}$$
  
  In a similar way the r.h.side $B^{n-1}_{min}$ can be  substituted by $$\frac{n-3}{2}B^{n-1}_{(j_{1}=\frac{n-1}{2},  j_2= \frac{n-3}{2})}$$
So, by taking into account the definitions of the splitting sequences and Green's functions in $\Phi_0$ \ref{def.2.4}, \ref{def.2.5}, as well as proposition \ref{prop.2.2}
(upper bounds of tree terms $C^{n+1}$ we require in a equivalent way (after some elementary simplifications) the sufficient condition:
\begin{equation}
\begin{array}{l}
\displaystyle{\frac{4\delta_{\frac{n+5}{2},min}{\mathcal T}_{\frac{n+5}{2}}(n-2)}{\delta_{n,min}{\mathcal T}_{n}'(n+1)}}>\ 1\\
\ \ \\
\mbox{or using the definitions of $\delta's$ and ${\mathcal T}_{n}'s$ we require the following  stronger condition:}\\
\ \ \\
f_B= \frac{(n-2)(n+5)(n+3)[1+3\Lambda n(n-1)][(n-1)^2+64(n-1)+192]}{(n+1)[4+3\Lambda (n+5)(n+3)]n(n-1)[(n-3)^2+16(n-3)+48]}>\ 1 
\end{array}
\label{5.97}
\end{equation}
The difference between the numerator and the denominator of $f_B $, let us say $Num_{fB}- Den_{fB}$, is a polynomial with positive coefficients of the dominant contributions:
 $$Num_{fB}- Den_{fB}\ \sim ({\cal O}(n^7)+{\cal O}(n^6)+{\cal O}(n^5))>0    $$
 Another way to be convinced that  the condition \ref{5.97} is verified is to put $n= x$ (continuous) and represent graphically the function $f_B(x)$ at fixed $\Lambda=0.05$. The figure \ref{fig.1.2} shows precisely that $f_B$ decreases continuously always from larger values than $1$  up to the limit value of $1$.
 This ends the proof of proposition \ref{prop.3.3}\ \ \ $\blacksquare$
\end{itemize}

 \begin{figure}[h]
\begin{center}
\hspace*{-5mm}
 \includegraphics[width=10cm]{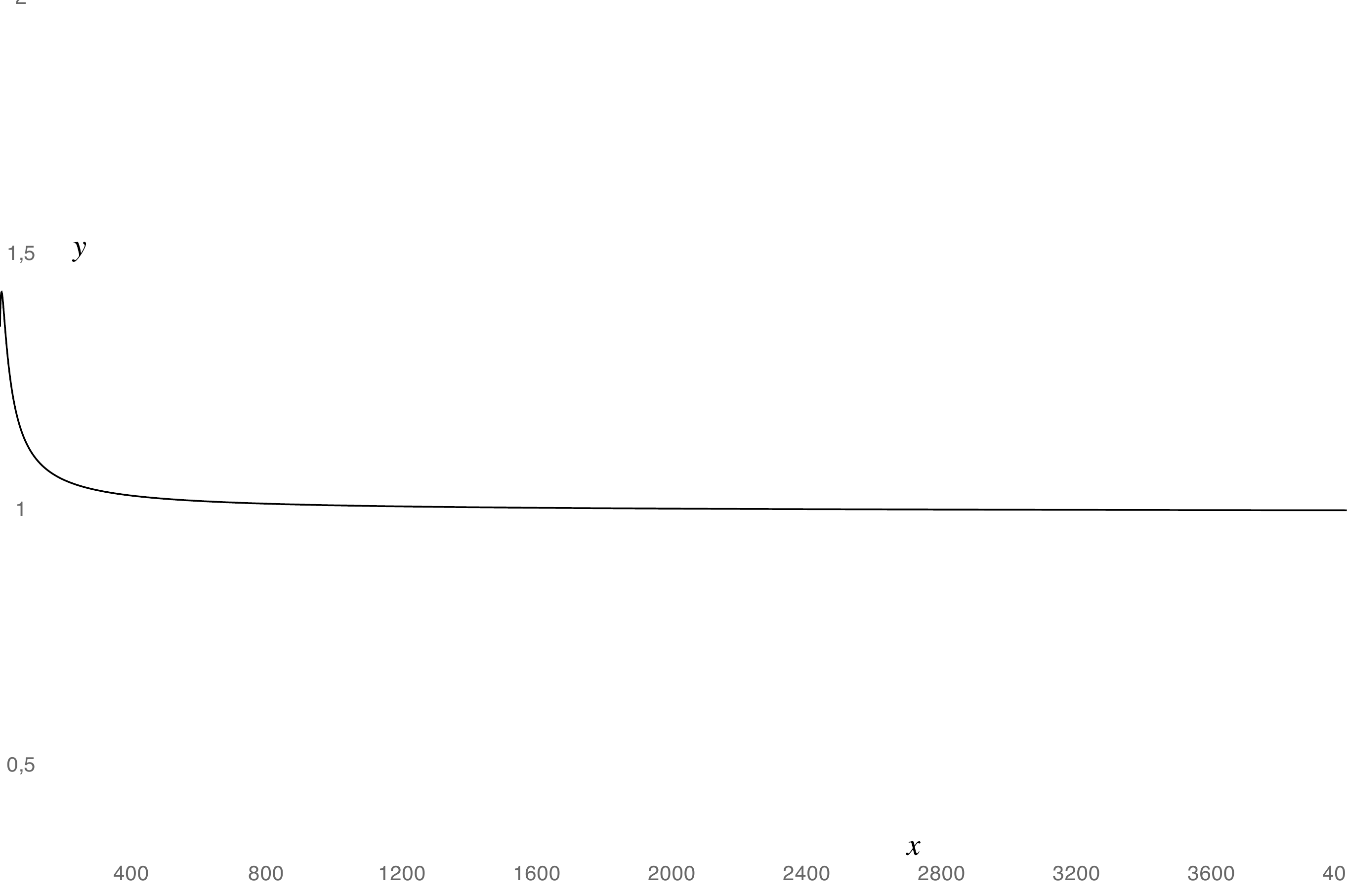}
\end{center} 
 \caption{\small\textrm{For the values of $n$ ($=x$ continuous) in the interval $]7, 4001] $ $f_B$ decreases continuously always from bigger values than $1$  up to the limit value of $1$}}
\label{fig.1.2}
\end{figure}

 

\vfill\eject   
 \item{\textbf{Proof of proposition \ref{prop.3.4}}}
  \begin{description}
  \item{i)} Using the hypothesis $H  \in \Phi$ we obtain the limit: $$\displaystyle{\lim_{\Lambda\rightarrow 0}D_n(\Lambda)}=0\quad \forall  n \geq 7$$
\item{ii)}The increasing property is a direct consequence of the previous propositions \ref{prop.3.2}, \ref{prop.3.3}.
\item{iii)} The lower and upper bounds together with the limit at infinity are also consequences of the same propositions.
\end{description}
\end{enumerate}

\vspace{3mm}

\subsection{Proof of theorem \ref{Th.3.1}}

\begin{description}
\vspace{3mm}
\item{i.} \ref{1.2.10}  For $n=1, 3, 5$: the limit properties (with respect to $\Lambda$) of the $H^2, H^4, H^6$ functions are ensured 
in view of the hypothesis $H\in\Phi$. The signs and bounds are established in proposition \ref{prop.3.1}
\item{ii.}  For
$n\geq 7$ propositions \ref{prop.3.2}, \ref{prop.3.3} and \ref{prop.3.4} yield the simple and uniform  convergence at infinity together  with the uniform bounds and limit properties (with respect to $\Lambda$). 

In particular the lower uniform  bound $d_0$ of $D_n's$ ensures the positivity of $\delta's$ and the the upper bound at infinity $\delta_{\infty}$ of $\delta_{n,max}$. Automatically the good signs and bounds are satisfied.
 $$\mbox{Moreover:}\qquad\lim_{\Lambda\rightarrow0}\, \frac{\delta_n^{'}
(\Lambda)}{\Lambda}=
 \lim_{\Lambda\rightarrow 0}\,\frac{3n(n-1)}{1+D_n (H)}$$
In view of the hypothesis $H\in \Phi$ we apply
 the property
 $\displaystyle{\lim_{\Lambda\rightarrow 0}D_n(\Lambda)}=0$
 of proposition \ref{prop.3.4} and the limit at $\Lambda=0$
is obtained.
 \end{description}

\section{Appendix II}\label{Ap.2}
{\bf Proof of theorem \ref{Th.4.1}}\ \
\subsection{The completeness of $\Phi_0$}
\begin{description}
\item[i.]  The ball  $S_{\rho}(H_0)\subset \Phi_0$ is by definition a closed subspace of ${\cal B}_c $ which is a complete metric space with respect to the topology defined in terms of the norm ${\cal N}$ (cf. \ref{def.4.1}). Consequently $S_{\rho}(H_0)$ is also complete with respect to the induced topology.  
\ \ $\blacksquare$.

In the \emph{numerical study} of \cite{MMST} we have precisely realized
this last result:

The solution is obtained iteratively in the neighborhood of this
 particular  sequence $H_0$ by using two starting points: the sequences $\{\delta_{n,max}\}$ and  $\{\delta_{n,min}\}$.
\end{description}
Let us now present the proof of the analogous result ``theoretically''.


\subsection{The local contractivity }

In this part of the paper we  have to show that for every  $H$   in $S_{\rho}(H_0)$  and when 
 $\Lambda\in ]0, 0.45]$ 
 there exist two real positive constants  (continuous
 functions of $\Lambda$), $k^{(0)}< 1$ and $k<1$ with
   $$k^{(0)}(\Lambda)<  (1-k(\Lambda))$$ 
such that:
\begin{equation}
\begin{array}{l}
 \Vert\mathcal{M}^{*}(H)-H_{(0)}\Vert\ \leq \ k(\Lambda)\ \rho\\ \mbox{and}\\
 \Vert\mathcal{M}^{*}(H_0)-H_{(0)}\Vert\ \leq \ k^{(0)}(\Lambda)\ \rho\
 \end{array}
     \label{6.90}
\end{equation}
 By 
 the definition \ref{def.4.1}
   \ of the norm $\mathcal{N}$
the above inequalities are equivalent to the following:
\begin{equation}
\begin{array}{l}
\forall\ \ \ \Lambda \in ]0, 0.45]\\
   \qquad \qquad\displaystyle{\sup_{(n,\Lambda)}}
\displaystyle{\frac{\vert H^{n+1^\prime}-H_{(0)}^{n+1}\vert}
    {M_{n}} }
   \leq \ k(\Lambda)\  \rho\\
   \mbox{and}\ \ \ \qquad\qquad\\
    \qquad\qquad \displaystyle{\sup_{(n,\Lambda)}}
\displaystyle{\frac{\vert H_{(0)}^{n+1^\prime}-H_{(0)}^{n+1}\vert}
    {M_{n}} }
   \leq \  k^{(0)}(\Lambda)\  \rho
     \end{array}
\label{6.91}
\end{equation}

By using definitions \ref{def.2.6} (of the sequence $\{H_0\}$), \ref{def.4.1}
   \ (of the norm $\mathcal{N}$)  and for $H\in S_{\rho}$, we obtain successively:
\begin{enumerate}
\item{}  
 for $n=1$  
$$
\vert H^{2^\prime}(\Lambda)-H_{(0)}^{2}(\Lambda)
\vert\leq
\Lambda \vert H^{4}-H_{(min)}^{4}\vert\leq
\Lambda[\vert H^{4}-H_{(0)}^{4}\vert+
\vert H_{(0)}^{4}-H_{(min)}^{4}\vert]
$$
\begin{equation}
    \leq 12\Lambda^{2}M_1^{3}\rho 
        \label{4.5207}
\end{equation}
and, we  write~:
\begin{equation}
    \frac{\vert H^{2^\prime}(\Lambda)-H_{(0)}^{2}(\Lambda)\vert}
    {M_{1}(\Lambda)}\leq k_{1}(\Lambda)\rho
    \label{4.5.208}
\end{equation}
where we define~:
\begin{equation}
k_{1}(\Lambda)=12\Lambda^{2}M_1^{2}
\label{6.99}
\end{equation}

Moreover:
\begin{equation}
\vert H_{(0)}^{2^\prime}(\Lambda)-H_{(0)}^{2}(\Lambda)
\vert\leq
\Lambda \vert H_{(0)}^{4}-H_{(min)}^{4}\vert \leq 6\Lambda^{2}M_1^{3}\rho 
 \label{4.5.207}
\end{equation}
   and, we  write~:
\begin{equation}
    \frac{\vert H(0)^{2^\prime}(\Lambda)-H_{(0)}^{2}(\Lambda)\vert}
    {M_{1}(\Lambda)}\leq k_{1}^{(0)}(\Lambda)\rho
    \label{4.5.208}
\end{equation}
where we define~:
\begin{equation}
k_{1}^{(0)}(\Lambda)=6\Lambda^{2}M_1^{2}
\label{6.102}
\end{equation}
We verify that 
\begin{equation}
k_{1}^{(0)}(\Lambda)+k_{1}(\Lambda) < 1 \ \ \ \forall \  \Lambda\leq 0.1 
\label{6.103}
\end{equation}

 \item{}
 For $n=3$ (in view of the definition \ref{def.3.1} of ${\cal M}^* $) we have :
\begin{equation}
    \vert H^{4{}^\prime}(\Lambda)-H_{(0)}^{4{}}
(\Lambda)\vert
    \leq \vert\delta_{3}'-\delta_{3(0)} \vert
(H^{2{}^\prime})^3+\delta_{3(0)}\vert
(H^{2{}^\prime})^3-(H_{(0)}^{2})^3\vert 
\label{6.104}  
\end{equation}
and,
\begin{equation}
\vert\delta_{3}'-\delta_{3(0)} \vert\leq 
\frac{6\Lambda\vert D_{3}-D_{3(min)}\vert }
{(1+D_{3})(1+D_{3(min)})}
\label{4.5.212} 
\end{equation} 
Here, from definition \ref{def.2.6}:
$$D_{3(min)}= 9\Lambda H^{2}_{min}
 - \displaystyle{\frac{\Lambda
|H^6_{min}|}{ |H^4_{max}|}. }$$
So, 
\begin{equation}
\vert\delta_{3}'-\delta_{3(0)} \vert\leq 
\frac{54\Lambda^2\{| H^{2}- H^{2}_{min}|
 - \Lambda(\frac{ 
| H^6|}{ |H^4|} -\  \frac{ |H^6_{min}|}{ |H^4_{max}|})\} }{(1+D_{3})(1+D_{3(min)})}
\end{equation}
Now in view of the bounds and sign properties of the Green's functions inside $\Phi_0$, the second term on the numerator is positive and can be eliminated, precisely:
\begin{equation} (\frac{ 
| H^6|}{ |H^4|} -\  \frac{|H^6_{min}|}{ |H^4_{max}|})=
 \displaystyle{\vert\frac{ 
 | H^6| }{ | H^4| } -\  \frac{|H^6_{min}|}{| H^4_{max}|}\vert}
 \label{6.107}
\end{equation}
Moreover:
$$\frac{6\Lambda}{1+D_{3}}\leq \delta _{3,max}\quad \mbox{and}\quad  \frac{6\Lambda}{1+D_{3(min)}}\leq \delta _{3,max}.$$
Finally the first term on the right hand sight (r. h.s.) of \ref{6.104}  yields:
\begin{equation}
\vert\delta_{3}'-\delta_{3(0)} \vert
(H^{2{}^\prime})^3 \leq
3M_1^{4}(\delta _{3,max})^2\rho
\label{4.5.213} 
\end{equation}
Taking into account the previous results \ref{4.5.208}, \ref{6.102} for $n=1$, the second term on the r.h.s. of \ref{4.5.211} yields: 
\begin{equation}
\delta_{3(0)}\vert
(H^{2{}^\prime})^3-(H_{(0)}^{2})^3\vert \leq 3 k_1 (\Lambda^2)M_1^{3}\delta_{3(max)}
\label{4.5.211}  
\end{equation}
From the above results and the norm definition we have:
\begin{equation}
   \displaystyle{\frac{\vert H^{4{}^\prime}(\Lambda)-H_{(0)}^{4{}}|}{M_3}} \leq  k_3(\Lambda)\rho
    \end {equation}
with:
\begin{equation}
k_3(\Lambda)=9\Lambda M_1(1+4\Lambda)
\label{6.111}
\end{equation}
In an  analogous way we obtain: 
\begin{equation}
   \displaystyle{\frac{\vert H_{(0)}^{4{}^\prime}-H_{(0)}^{4{}}|}{M_3}} \leq  k_3^{(0)}(\Lambda)\rho
    \end {equation}
with:
\begin{equation}
k_3^{(0)}(\Lambda)=9\Lambda \displaystyle{\frac{(H_{(0)}^2)^2(1+2\Lambda)}{M_1^2}}
\label{6.113}
\end{equation}
We finally verify that:
\begin{equation}
\forall\ \  0<\Lambda\leq 0.045 \qquad k_3(\Lambda)+k_3^{(0)}(\Lambda)< 1
\label{6.114}
\end{equation}

 \item{}
 For $n=5$ using the definition \ref{def.3.1} of the mapping we have:
\begin{equation}
    \vert H^{6{}^\prime}(\Lambda)-H_{(0)}^{6{}}
(\Lambda)\vert
    \leq \displaystyle{\frac{\vert\delta_{5}'-\delta_{5(0)} \vert
\vert C^{6^{}\prime}\vert}{60\Lambda}}+\displaystyle{\frac{\delta_{5(0)}\vert
C^{6^{}\prime}-C_{(0)}^{6^{}}\vert}{60\Lambda}.}
\label{6.115}  
\end{equation}
In view of the stability of $\Phi_0$, definitions \ref{def.2.6} and \ref{def.4.1} (of the norm), the first term of the r.h.s. of
the latter yields: 
\begin{equation}
\begin{array}{c}
\displaystyle{\frac{\vert\delta_{5}'-\delta_{5(0)} \vert
\vert C^{6^{}\prime}\vert}{60\Lambda}}\leq \vert\delta_{5(max)}-\delta_{5(min)} \vert
\vert H_{(max)}^{4{}}\vert (H_{(max)}^{2{}})^2\\
\leq \delta_{5(max)}M_3M_1^{2}\rho
\end{array}
\label{6.116} 
\end{equation} 
By the earlier results for $n=1\  \mbox{and}\ \ n=3$, the second term of the r.h.s. of \ref{6.115} reads:
\begin{equation}
\displaystyle{\frac{\delta_{5(0)}\vert
C^{6^{}\prime}-C_{(0)}^{6{}}\vert}{60\Lambda}} 
\leq \delta_{5(max)}M_3M_1^{2}\rho [k_3(\Lambda)+2 k_1(\Lambda^2)]
\label{6.117}
\end{equation}
Then, from \ref{6.116}, \ref{6.117} and the norm definition\ \ref{def.4.1} ($M_5=20M_3M_1^2 \delta_{5(max)}$), we have:
\begin{equation}
\begin{array}{l}
    \vert H^{6{}^\prime}(\Lambda)-H_{(0)}^{6{}}
(\Lambda)\vert \leq \rho\  k_5(\Lambda)\\
 \mbox{ where:} \quad k_5(\Lambda)= \displaystyle{\frac{1}{20}(1+k_3(\Lambda)+ 2k_1(\Lambda^2))}
    \end{array}
    \end{equation}
In an analogous way we find:
\begin{equation}
\begin{array}{l}
    \vert H^{6{(0)}^\prime}(\Lambda)-H_{(0)}^{6{}}
(\Lambda)\vert \leq \rho\  k_5^{(0)}(\Lambda)\\
 \mbox{ where:} \quad k_5^{0)}(\Lambda)= \displaystyle{\frac{1}{20}(1+k_3^{(0)}(\Lambda)+ 2k_1^{(0)}(\Lambda^2))}
    \end{array}
    \end{equation}
    and, verify that under the same condition on the coupling constant as  that imposed previously in the case $n=3$ we obtain:
\begin{equation}
\forall\ \  0<\Lambda\leq 0.045 \qquad k_5(\Lambda)+k_5^{(0)}(\Lambda)=0,25 < 1
\label{6.120}
\end{equation}
consequently for $n=5$ we require a weaker condition on $\Lambda.\ \ \qquad \ \qquad \blacksquare$. \ \ \


 \item{}

 For $n\geq 7$, we proceed by recursion~:
 
 \textbf{{h.c.r}}: We suppose that for every
$m\leq n-2$ there exist two finite positive  constants (continuous functions of $\Lambda$) $k_{m}(\Lambda)$ and  $k_{m}^{(0)}(\Lambda)$
 precisely given by: 
\begin{equation}
\begin{array}{l}
k_{m}(\Lambda)=a_1\displaystyle{\sum_{j=0}^{\frac{m-5}{2}}(\frac{1}{12})^j +  (\frac{1}{(12)})^{\frac{m-5}{2}}}k_5;\\ 
k_{m}^{(0)}(\Lambda)=a_0\displaystyle{\sum_{j=0}^{\frac{m-5}{2}}(\frac{1}{12})^j +(\frac{1}{12})^{\frac{m-5}{2}}k_5^{(0)}}\\
\mbox{with:}\quad  a_0=\displaystyle{\frac{1}{12}+\frac{k_1^{(0)}(\Lambda^2)}{6}}\ \quad \mbox{and} \quad  a_1=\displaystyle{\frac{1}{12}+\frac{k_1(\Lambda^2)}{6}}
\end{array}
\label{6.121}
\end{equation}
 such that:
\begin{equation}
\begin{array}{l}\qquad\qquad
\forall\  \Lambda \in ]0, 0.45]\\
\ \ \\
   \qquad \qquad
\displaystyle{\frac{\vert H^{m+1^\prime}-H_{(0)}^{m+1}\vert}
    {M_{n}} }
   \leq \ k_{m}(\Lambda)\  \rho\\
   \mbox{and},\ \ \ \qquad\\
    \qquad\qquad \displaystyle{\frac{\vert H_{(0)}^{m+1^\prime}-H_{(0)}^{m+1}\vert}
    {M_{n}} }
   \leq \  k_m^{(0)}(\Lambda)\  \rho\\
  \mbox{ with:} \ \ \qquad\ \\
   \qquad  \qquad k_{m}(\Lambda)+ k_m^{(0)}(\Lambda)\ <\ 1
     \end{array}
\label{6.122}
\end{equation}

We establish the statement \textbf{\ {h.c.r}} for $m=n$  by using  arguments analogous to that for $m=5$:

By the definition \ref{def.3.1} of the mapping we have:
\begin{equation}
    \vert H^{n+1{}^\prime}-H_{(0)}^{n+1{}}
(\Lambda)\vert
    \leq \displaystyle{\frac{\vert\delta_{n}'-\delta_{n(0)} \vert
\vert C^{n+1^{}\prime}\vert}{3\Lambda n(n-1)}}+\displaystyle{\frac{\delta_{n(0)}\vert
C^{n+1^{}\prime}-C_{(0)}^{n+1^{}}\vert}{3\Lambda n(n-1)}.}
\label{6.123}  
\end{equation}
In view of the stability of $\Phi_0$, definitions \ref{def.2.6} and \ref{def.4.1} (of the norm), the first term of the r.h.s. of
the latter yields: 
\begin{equation}
\begin{array}{c}
\displaystyle{\frac{\vert\delta_{n}'-\delta_{n(0)} \vert
\vert C^{n+1^{}\prime}\vert}{3\Lambda n(n-1)}}\leq \vert\delta_{n(max)}-\delta_{n(min)} \vert \ \mathcal{T}_{n}\ 
\vert H_{(max)}^{n-1{}}\vert (H_{(max)}^{2{}})^2\\
\leq \delta_{n(max)}\ {\mathcal{T}_{n}}\ M_{n-2}M_1^{2}\rho
\end{array}
\label{6.124} 
\end{equation} 
Here we have taken into account the bound of the tree term $C^{n+1^{}\prime}$ given by Proposition \ref{prop.2.2} in terms of the dominant contribution $\vert H_{(max)}^{n-1{}}\vert (H_{(max)}^{2{}})^2$.

$\mathcal{T}_{n}$ means the number 
 of different partitions
in the sum $\sum\limits_{\varpi_{n}(I)}$ of the tree term 
(cf. App. A of reference \cite[c]{MM1}) which  
 verifies :\\
\begin{equation}
 \forall\ \ \  n>9,  \qquad 
\mathcal{T}_{n}=\frac{(n-3)^{2}}{48}+\frac{(n-3)}{3}+1,
    \label{6.125}
\end{equation}
\begin{guess} \emph{We remind that   $\mathcal{T}_{7}=2 $ and $\mathcal{T}_{9}=3 $ because only the two last terms in the expression \ref{6.125} yield a nontrivial contribution.} 
\end{guess}

By the recurrence hypothesis for $n-2\  \mbox{and the earlier result for }\ \ n=1$, the second term of the r.h.s. of \ref{6.123} (tree term) yields:
\begin{equation}
\displaystyle{\frac{\delta_{n(0)}\vert
C^{n+1^{}\prime}-C_{(0)}^{n+1^{}}\vert}{3\Lambda n(n-1)}}\leq \delta_{n(max)}Ö \mathcal{T}_{n}\ M_{n-2}M_1^{2}\rho [k_{n-2}(\Lambda)+2 k_1(\Lambda^2)]
\label{6.126}
\end{equation}
Then, from \ref{6.124}, \ref{6.125}, \ref{6.126} the norm definition \ref{def.4.1}: $$M_n=n(n-1)M_{n-2}M_1^2 \delta_{n(max)},$$ and the bound:
$$\displaystyle{\frac{1}{48n(n-1)}[(n-3)^2]+16(n-3)+48]<\ \frac{4}{48}=\frac{1}{12}} \ \qquad \forall n\geq 7)$$
 we obtain:
\begin{equation}
\begin{array}{l}
    \vert H^{n+1{}^\prime}(\Lambda) - H_{(0)}^{n+1{}}
(\Lambda)\vert \leq \rho \ k_n(\Lambda)\\
 \mbox{ where:} \quad k_n(\Lambda)= \displaystyle{ \frac{1}{12}+\frac{k_{n-2}}{12}+\frac{ k_1(\Lambda^2)}{6} } 
    \end{array}
    \end{equation}
    
In an analogous way we find:
\begin{equation}
\begin{array}{l}
    \vert H_{(0)}^{n+1{}^\prime}(\Lambda) - H_{(0)}^{n+1{}}
(\Lambda)\vert \leq \rho\  k_n^{(0)}(\Lambda)\\
 \mbox{ where:} \quad k_n^{(0)}(\Lambda)= \displaystyle{ \frac{1}{12}+\frac{k_{n-2}^{(0)}}{12}+\frac{ k_1{(0)}(\Lambda^2)}{6}} 
    \end{array}
    \end{equation}
Now, by the expression of he recursion statement \textbf{{h.c.r}} \ref{6.121} for  $k_{n-2}$ and respectively for $k_{n-2}^{(0)}$ we obtain:
\begin{equation}
\begin{array}{l}
k_n(\Lambda)= \displaystyle{ \frac{1}{12}+\frac{1}{12}a_1\displaystyle{\sum_{j=0}^{\frac{n-7}{2}}(\frac{1}{12})^j +(\frac{1}{12})^{\frac{n-7}{2}}k_5}+\frac{ k_1(\Lambda^2)}{6} }=\\ 
= a_1\displaystyle{\sum_{j=0}^{\frac{n-5}{2}}(\frac{1}{12})^j +(\frac{1}{12})^{\frac{n-5}{2}}k_5}\\
\mbox{and  respectively,}\ \ \\
k_n^{(0)}(\Lambda) = a_0\displaystyle{\sum_{j=0}^{\frac{n-5}{2}}(\frac{1}{12})^j +(\frac{1}{12})^{\frac{n-5}{2}}k_5^{(0)}}
\end{array}
\label{6.129}
\end{equation}
We directly verify that:
\begin{equation}
\forall \ \ \ \Lambda \in ]0, 0.45]\quad k_{n}(\Lambda)+ k_n^{(0)}(\Lambda)\ <  1
\end{equation}
This result ends the proof of the recursive formula\ \ \ \ $\qquad \qquad\blacksquare$

Finally, taking into account the limit of the convergent geometric series in the final results \ref{6.129} we also verify that: 

\begin{equation}
\begin{array}{l}
\displaystyle{\sup_{n} \ k_{n}(\Lambda)=\ \lim_{n\to\infty}\{\ a_1\sum_{j=0}^{\frac{n-5}{2}}(\frac{1}{12})^j +(\frac{1}{12})^{\frac{n-5}{2}}k_5}\}=\\
=\frac{1}{11}+\frac{ k_1(\Lambda^2)}{6}
\end{array}
\end{equation}
and by analogy:
\begin{equation}
\begin{array}{l}
\displaystyle{\sup_{n} \ k_{n}^{(0)}(\Lambda)=\ \lim_{n\to\infty}\{\ a_1\sum_{j=0}^{\frac{n-5}{2}}(\frac{1}{12})^j +(\frac{1}{12})^{\frac{n-5}{2}}k_5}\}=\\
=\frac{1}{11}+\frac{ k_1^{(0)}(\Lambda^2)}{6}
\end{array}
\end{equation}
Then we introduce:
\begin{equation}
\begin{array}{l}
k=\displaystyle{\sup_{(n;\Lambda\in]0,0.045])}(k_{n}(\Lambda)}\\
\mbox{and in a similar way}\\
k^{(0)} =\displaystyle{\sup_{(n;\Lambda )}k_{n}^{(0)}(\Lambda)} 
\end{array}
\end{equation}
By the definitions \ref{6.99}, \ref{6.102} of $k_1(\Lambda^2)$
and $k_1^{(0)}(\Lambda^2)$ we find:
\begin{equation}
k=\frac{1}{11}+0.005=0.096;\ \ \ \mbox{and}\ \ k^{(0)}=0.094
\end{equation}

\textbf{Conclusion:}
\begin{equation}
\begin{array}{l}
\forall\ \ \ \Lambda \in ]0, 0.45]\\
   \qquad \qquad\displaystyle{\sup_{(n,\Lambda)}}
\displaystyle{\frac{\vert H^{n+1^\prime}-H_{(0)}^{n+1}\vert}
    {M_{n}} }
   \leq \ k\  \rho\\
   \mbox{and}\ \ \ \qquad\qquad\\
    \qquad\qquad \displaystyle{\sup_{(n,\Lambda)}}
\displaystyle{\frac{\vert H_{(0)}^{n+1^\prime}-H_{(0)}^{n+1}\vert}
    {M_{n}} }
   \leq \  k^{(0)} \  \rho  \\
   \mbox{with:} k+ k^{(0)} < 1\qquad \qquad\ \ \blacksquare
     \end{array}
\label{6.91}
\end{equation}

\end{enumerate}



\begin{thebibliography}{99}

\bibitem{(Q.F.T.)}
    \begin{itemize}
\item[a)] A.S. Wightman Phys. Rev. 101, 860 (1965)

    \item[b)] R. Streater and A. Wightman.
\emph{PCT Spin Stat.and all That} (Benjamin, New York,1964)

    \item[c)] N.N. Bogoliubov, A.A. Logunov, and I.T. Todorov.
 \emph{Introduction to the Axiomatic Q.F.T.}

    (Benjamin, New York, 1975)

        \item[d)] R. Jost.\emph{The General Theory
of Quantized Fields}
         (American Math.Society,Providence,RI,1965)

                                    \item[e)] N.N. Bogoliubov, D.V. Shirkov,
\emph{Introduction to the Theory of Quantized Fields}

                                    (Interscience, New York, 1968)
\end{itemize}

\bibitem{MM1}M. Manolessou
\begin{itemize}
\item[a)] J. Math. Phys. 20  2092 (1988)
    \item   [b)] 30 175 (1989)
        \item [c)]  30  907 (1989)
\item[d)] J. Math. Phys.32 12 (1991)
    \item[e)] \emph{Back to the  $\Phi^4_0$  solution}
 Preprint  E.I.S.T.I.(1994)
\end{itemize}

    \bibitem{MM2}M. Manolessou
\begin{itemize}
\item[a)] Nucl. Physics B (Proc. Suppl.) 6 (1989) 163-166
North-Holland \item[b)]\emph{The  $\Phi^4_4$ non trivial solution}
 Preprint  E.I.S.T.I.(1992)
\item[c)] Contribution to the $XI^{th}$
                                        International Congress of Math.Physics Unesco-Sorbonne
                                        (D.Iagolnitzer editor 1994)
\end{itemize}

\bibitem{G.J.}J. Glimm and A. Jaffe
\begin{itemize}
\item[a)] Phys. Rev. 176, 1945 (1968); \item[b)] Commun. Math.
Phys.11,99 (1968); 
\item[c)] Bull.Ann Math.Soc;76 407 (1969);
\item[d)] Acta. Math.125, 203 (1970); \item[e)]\emph{Stat.Mech.
and Quantum Field Theory}
                            Les Houches,1970(1-108) (Gordon and Breach, N.York, 1971).
\end{itemize}
\bibitem{Sym} K.  Symanzik J.Math.Phys. 7, 510 (1966)
\bibitem{G.J.1} J. Glimm and A.  Jaffe,
                                                            Commun. Math. Phys. 22 253, (1971)
\bibitem{G.J.S.} J. Glimm and A. Jaffe, and T. Spencer.
                                                            \emph{Constructive Quantum Field Theory},
                                                         Lecture Notes in Phys.Vol.25
                                                G.Velo and A. Wightman (Springer, 1973)

\bibitem{Zim1}W. Zimmermann\
\begin{itemize}
\item[a)]Commun.Math.Phys. 6,161 (1967) 
\item[b)]10, 325 (1968)
\end{itemize}
\bibitem{MM3}M. Manolessou, Ann.Phys.(NY)152, 327 (1984)
\bibitem{Dys}F.J. Dyson, Phys.Rev.75, 486, 1736 (1949)
\bibitem{Sch}J.Schwinger, Phys.Rev.75, 651,76 (1949)
\bibitem{Br.MM}J. Bros-M. Manolessou-Grammaticou,
Commun.Math.Phys.72(1980)175-205,207-237
\bibitem{MM4}M. Manolessou-Grammaticou, Ann.Phys.(NY)122,(1979)
\bibitem{MMST} M. Manolessou-S. Tafat 
``Numerical study of the local contractivity of the $\Phi^4_0$ mapping'' Preprint E.I.S.T.I September 2011
\bibitem{MM5} M. Manolessou The $\Phi_4^4$ nontrivial solution. 
Preprint E.I.S.T.I. in preparation
\end{thebibliography}
\end{document}